\documentstyle[aps,eqsecnum]{revtex}
\begin{document}
\title{\hspace{12.0cm} Preprint WSU-NP-96-13 \protect\\
\vskip 1.5cm
The wedge form of relativistic dynamics. II. The Gluons }            
\author{ A. Makhlin}
\address{Department of Physics and Astronomy, Wayne State University, 
Detroit, MI 48202}
\date{\today}
\maketitle
\begin{abstract}
I derive  expressions for  various correlators of the gauge  field and
find propagators in a new gauge $A^\tau=0$. This gauge is a part of the
wedge form of relativistic dynamics suggested earlier \cite{WD1} as the
tool for the study of quantum dynamics in collisions of hadrons at
extremely high energies and in ultrarelativistic heavy ion collisions. 
The new gauge  puts the quark and gluon fields of the
colliding hadrons in one Hilbert space and thus allows one to avoid 
factorization.                                                         
\end{abstract}

\section{Introduction}\label{sec:SN1}

In the previous paper \cite{QGD} I  explained physical motivation
of the ``wedge form of  dynamics'' as a promising tool to explore the
processes which take place during the collision of two heavy ions. In
compliance with the general definition of dynamics given by  Dirac
\cite{Dirac}, the wedge form  includes its specific definition of  the
quantum mechanical observables on the  space-like surfaces, as well as the
means to describe evolution of the observables  from the ``earlier''
space-like surface to the ``later'' one.  Unlike the other forms,
the wedge form explicitly refers to the two main geometrical features
of the phenomenon, strong localization of the initial
interaction and, as a consequence, the absence of the translational
invariance in the temporal and longitudinal directions.

Usually, every hadron (or nucleus)  before the collision are considered
separately, in their own infinite momentum frame. Thus, we always deal
with
two different Hamiltonian dynamics with their own definition of the time
variable. This problem is seemingly resolved by the factorization scheme
which replaces the true bound state by an artificial flux of free partons.
However, the class of the physical processes which comply with this
method is restricted to the inclusive production of jets with the high
transverse momenta. Factorization is obviously violated already in the
diffractive deep inelastic electron-proton scattering and, in fact, is
not the case in any semi-inclusive process.  

The constraints imposed by the factorization become critical when one is
compelled to consider interaction of the two bounded systems  without
appealing to the parton picture \cite{QGD}, which is the case of the
heavy-ion  collision. One cannot design this experiment in such a way that
the scale of the ``hard probe,'' like, {\em e.g.}, the dilepton mass in
the Drell-Yan process, will become a measured parameter. In fact, the
nuclei {\em a priori} probe each other at all scales and the expected
factorization scale turns into ill-defined infra-red cut-off.  Therefore,
it is necessary to find a  way to describe both colliding systems, the
hadrons or  the nuclei, using {\it the same Hamiltonian dynamic.}  This
requirement follows solely from the fact that the definition of the field
states (particles) depends on how the observables are defined.  The
problem is most painful for gluons since the  choice of the  gauge is one
of the elements of the Hamiltonian dynamic.  It manifestly affects the
definition of the physical states of the gauge fields. Indeed, since every
dynamics has a specific definition of the time variable,  Gauss' law which
{\em  defines}  the non-dynamical longitudinal fields, also looks
differently. 

Factorization succeeds to fragment the cross section of the hadronic
process into the  hard partonic cross section, which is gauge invariant
because the partons are considered as the  free on mass shell particles, 
and the gauge invariant structure functions.  By the failure of the
factorization, one is lead back to the generic field-theory approach. It
strongly requires  the  gluon fields from the incoming hadrons be given
within the same gauge.

The gluons are expected to be most abundant and active at the early stage 
of the heavy-ion collision. According to estimates of Shuryak and Xiong
\cite{Shur} most of the entropy should be produced due to the gluon
interactions. However, before the collision, the quark and the gluon
fields are assembled into the two coherent wave packets, the nuclei, and,
therefore, the initial entropy equals zero. The coherence is lost and the
entropy is created only due to interaction. Though it is tempting to rely
on the general formula, $S={\rm Sp}\rho\ln\rho$, which invariantly
expresses the entropy via the density matrix $\rho$, at least one basis of
states should be given explicitly. It is imperative to find what are the
states that may be used for computing the entropy.

In the wedge form of dynamics, the states of the quark and gluon fields
are defined on  the space-like hypersurfaces of the constant proper  time
$\tau$, $\tau^2=t^2-z^2$. The states of scalar and fermion fields were
discussed in Ref.\cite{WD1}. In this paper, we continue the study for the
gluons and augment our previous consideration by the gauge condition
$A^\tau =0$. This simple idea solves several problems. First, it becomes 
possible to treat two different light-front gauges  which describe  gluons
from each nucleus of the initial state separately, as  the two limits of
this single gauge. Therefore, the new approach keeps important 
connections with the theory of the deep inelastic e-p scattering. This
fact is vital for the subsequent calculations since the e-p deep inelastic
scattering is the only existing source of the data on the nucleon
structure in the high-energy collisions. [An alternative point of view is
based on the classical model of the large nucleus in the infinite momentum
frame \cite{McLer}.]  Second, after collision, this kind of a gauge 
becomes a  local temporal axial gauge , thus providing a smooth transition
to the Bjorken regime of the boost-invariant expansion.  

Most part of this paper is technical, and any relevant physical discussion 
of the results always goes only after their mathematical derivation. In
section \ref{subsec:SB21},  I derive equations of motion for the gauge
field in the gauge $A^\tau =0$, find Hamiltonian variables and
normalization condition. Equations of motion are linearized and the modes
of the free  radiation field are obtained in section \ref{subsec:SB22}. In
section \ref{subsec:SB23} the retarded propagator of the perturbation
theory is found as the response function of the field on the external
current.  I carefully examine possibility to separate dynamics of the
transverse and longitudinal fields and arrive at the negative conclusion. 
This part of calculation turned out to be the most time-consuming,
since the gauge condition is inhomogeneous and none of the currently used 
methods is effective. The old-fasioned variation of parameters does work.
To reassure its effectiveness, propagators of more familiar gauges,
$A^0=0$ and  $A^\pm=0$, were computed in Ref.~\cite{TAG}. In
section \ref{subsec:SB24} I show that the previously obtained propagator
solves the initial data problem for the gauge field. 

Section \ref{sec:SN3} is devoted to the quantization of the vector field
in the gauge $A^\tau=0$. I begin in \ref{subsec:SB31} with computation of
the Wightman functions and study the causal properties of the commutators
in \ref{subsec:SB32}. The latter appears to be abnormal, the Riemann
function is not symmetric and penetrate the exterior of the light cone.
However, behavior of the observables is fully causal and procedure of the
canonical quantization is accomplished in \ref{subsec:SB33}.  Even though
it is impossible to introduce transverse and longitudinal currents and
thus fully separate dynamics of the corresponding fields, I found useful
to classify various field patterns by the type of their propagation.
Propagator of the  transverse field is sensitive to the light cone
boundaries while  longitudinal and instantaneous parts of the field do not
really propagate. These two fragments of the response function are derived
in section \ref{sec:SN4}. In section \ref{sec:SN5} I study the limit
behavior of the propagator in the central rapidity region and in the
vicinity of the null-planes. I show that propagators of the gauges $A^0=0$
and  $A^\pm=0$, respectively are recovered.  This result is practically
important because it establishes connection of the new approach with the
existing theory of the deeply  inelastic processes at high energies.


\section{The classical treatment}
\label{sec:SN2}                

\subsection{Classical equations of motion}
\label{subsec:SB21}  

Here we consider the case of pure glue-dynamics.
We denote $A_\mu(x)=t^a A^{a}_{\mu}(x)$, the gluon field in the fundamental
representation of the color group. Consequently, we have the field tensor,
\begin{eqnarray}  
F_{\mu\nu}=t^a F^{a}_{\mu\nu}={\cal D}_\mu A_\nu-{\cal D}_\nu A_\mu=
\partial_\mu A_\nu -\partial_\nu A_\mu - ig [A_\mu,A_\nu], \nonumber
\end{eqnarray}
where ${\cal D}_\mu= \partial_\mu -ig[A_\mu(x), ...]$ is the covariant
derivative on the local color group.
The gauge invariant action of the theory looks as follows,
\begin{eqnarray}
{\cal S}=\int {\cal L}(x)d^4x =\int [-{1\over 4}
{\rm g}^{\mu\lambda}(x){\rm g}^{\nu\sigma}(x)F_{\mu\nu}(x)F_{\lambda\sigma}(x)
-j^\mu A_\mu ]\sqrt{-{\rm g}} d^4 x.
\label{eq:E2.1}\end{eqnarray}       
Its variation with respect to the gluon field yields the Lagrangian
equations of motion,
\begin{eqnarray}
\partial_\lambda [(-{\rm g})^{1/2}{\rm g}^{\mu\lambda}
{\rm g}^{\nu\sigma}F_{\mu\nu}] - 
ig (-{\rm g})^{1/2} [A_\lambda, {\rm g}^{\mu\lambda}
{\rm g}^{\nu\sigma}F_{\mu\nu}]=
(-{\rm g})^{1/2}j^\sigma~~,
\label{eq:E2.2}\end{eqnarray}
where $j^\mu$ is the color current of the fermion fields and $~~{\rm g}=
{\rm det}|{\rm g}_{\mu\nu}|$. The equations are twice covariant: with
respect to the gauge transformations in color space and the arbitrary
transformations of the coordinates. In what follows we shall employ the
special coordinates associated with the constant proper time  hypersurfaces
inside the light cone of the collision point $t=z=0$. The new coordinates
parameterize the Minkowsky coordinates $(t,x,y,z)$ as
$(\tau\cosh\eta,x,y,\tau\sinh\eta)$. In addition, we impose the gauge 
condition $A_\tau =0$.     The corresponding gauge transformation is well
defined.  Indeed, let $A_{\mu}(x)$ be an arbitrary field configuration and
$A'_{\mu}(x)$ its gauge transform with the generator 
\begin{eqnarray}
U(\tau,\eta, \vec{r_\bot})=
P_\tau \exp \{-\int_{0}^{\tau} A_\tau (\tau',\eta, \vec{r_\bot})d\tau'\}
\label{eq:E2.6}\end{eqnarray}   
Then the new field, $A'_{\mu}=U A_{\mu}U^{-1} + \partial_\mu UU^{-1}$, obey
the condition $A^\tau=0$.    
Imposing this gauge condition we arrive at the system of four equations:
\begin{eqnarray}
{\cal C}(x)={1\over \tau}\partial_\eta \partial_\tau A_\eta + \tau
\partial_r \partial_\tau A_r -ig\{{1\over\tau}[A_\eta,\partial_\tau A_\eta]
+\tau [A_r, \partial_\tau A_r]\} -\tau j^\tau = 0,
\label{eq:E2.3}\end{eqnarray}       
\begin{eqnarray}
-\partial_\tau \tau \partial_\tau A_r +
{1\over \tau}\partial_\eta(\partial_\eta A_r-\partial_r A_\eta)
 + \tau \partial_s (\partial_s A_r-\partial_r A_s) \nonumber \\
- ig\{{1\over \tau}\partial_\eta [A_\eta ,A_r] +\tau\partial_s[A_s,A_r]+
{1\over \tau}[A_\eta,F_{\eta r}] + \tau [A_s,F_{sr}]\} 
-\tau j^r =0~,
\label{eq:E2.4}\end{eqnarray}
\begin{eqnarray}
-\partial_\tau {1\over \tau} \partial_\tau A_\eta+ 
{1\over \tau}\partial_r (\partial_r A_\eta-\partial_\eta A_r)-
ig \big[{1\over \tau}\partial_r[A_r,A_\eta]
+{1\over\tau}[A_r,F_{r\eta}] \big]-\tau j^\eta =0
\label{eq:E2.5}\end{eqnarray}  
Here, we  use the latin indices from $r$ to $w$  for transverse $x$- and
$y$-components, $r,...,w=1,2$. We shall also use the arrows over the
letters to denote the  two-dimensional vectors, like ${\vec k}=(k_x,k_y)$,
$|{\vec k}|=k_{\bot}$. The latin indices from $i$ to $n$, $i,...,n=1,2,3$,
will be used  for the three-dimensional internal coordinates $u^i=
(x,y,\eta)$ on the hypersurface $\tau=const$. The metric tensor has only
diagonal components ${\rm g}_{\tau\tau}=-{\rm g}_{xx}= -{\rm g}_{yy}=1$,
${\rm g}_{\eta\eta}= -\tau^2$. The first of these equations contains no
second order time derivatives  and is a constraint rather than a dynamical
equation.  The constraint weakly equals to zero in classical Hamiltonian
dynamics and serves as the condition on physical states in quantum theory. 
The canonical momenta of the theory are as follows: 
\begin{eqnarray}
p^\tau=0,\;\;\;\; p^\eta={1\over \tau} F_{\tau\eta}= {1\over \tau}
\stackrel{\bullet}{A}_\eta,\;\;\;\;  
p_r=\tau F_{\tau r}=\tau\stackrel{\bullet}{A}_r~.
\label{eq:E2.7}\end{eqnarray} 
Hereon, the dot above the letter denotes derivative with respect to the
Hamiltonian time $\tau$.  Because of the gauge condition, the canonical
momenta do not contain the color commutators. After excluding the velocities,
the Hamiltonian can be written down  in the canonical variables: 
\begin{eqnarray} 
H=\int d\eta d \vec{r_\bot}
\tau\{{1\over 2} p^\eta p^\eta + {1\over 2\tau^2} p^r p^r +{1\over
2\tau^2} F_{\eta r}F_{\eta r}+ {1\over 4}F_{rs}F_{rs} +j^\eta A_\eta+ j^r
A_r\} 
\label{eq:E2.8}\end{eqnarray} 
Then the equations (\ref{eq:E2.4}) and (\ref{eq:E2.5}) are immediately
recognized as the Hamiltonian equations of motion. The Poison bracket of
the constraint ${\cal C}$ with the Hamiltonian vanishes thus creating the
generator of the residual gauge transformations which are tangent to the
hypersurface.  Conservation of the constraint is a direct consequence of
the Lagrange (or Hamiltonian) classical  equations of motion as well.

The normalization condition for the one-particle solutions is obviously 
derived from the charge conservation law. For the gauge field this is
impossible. Therefore we shall accept the condition which supports
self-adjointness of the homogeneous system after its linearization.
This leads to a natural definition for the scalar product of the states 
of the  vector field in the gauge $A^\tau =0$~:       
\begin{eqnarray}
(V,W)=\int_{-\infty}^{\infty}d\eta\int d^2{\vec r} \tau {\sf g}^{ik}
V^{*}_{i} i{\stackrel{\leftrightarrow}{\partial}}_{\tau} W_k
\label{eq:E2.14}\end{eqnarray}               
where ${\sf g}^{ik}$ is the metric tensor of the three-dimensional 
internal geometry of the hypersurface $\tau = const$~.  This norm of
the one-particle states prevents them from the flow out of the interior
of the past and future light wedges of the interaction plane.

\subsection{Modes of the free radiation field. Field of the static source}
\label{subsec:SB22}  

As a tool for the future development of the perturbation theory, we need
to find the propagators and Wightman functions when the nonlinear
self-interaction of the gluon field is
switched off. In this case the system of the equations for the
nonvanishing components of the vector potential  and the constraint look
as follows:
\begin{eqnarray}
[\partial_{\tau}\tau\partial_{\tau}-{1\over \tau }\partial_{\eta}^{2} 
-\tau\partial_{s}^{2}]A_r +\partial_{r}[\tau\partial_{s}A_s + 
{1 \over \tau}\partial_{\eta}A_\eta] =-\tau j^r~~,
\label{eq:E2.10}\end{eqnarray}         
\begin{eqnarray}
\big[\partial_{\tau}{1\over\tau} \partial_{\tau} -
{1\over\tau} \partial_{s}^{2}\big] A_\eta
+{1\over \tau}\partial_{\eta}\partial_{s}A_s =-\tau j^\eta~~,
\label{eq:E2.11}\end{eqnarray}
\begin{eqnarray}
{\cal C}(x)={1\over \tau}\partial_\eta \partial_{\tau}A_\eta+
\tau\partial_{\tau} \partial_{r}A_r -\tau j^\tau= 0~~.
\label{eq:E2.13}\end{eqnarray}
where $j^\mu$ includes all kinds of the color currents.            
An explicit form of the solution for the homogeneous system is found in
the Appendix 1. In compliance with the gauge condition which explicitly
eliminates  one of four field components we  find three modes
$V^{(\lambda)}$  of the free vector field. Two transverse modes obey Gauss
law without  the charge and have the unit norm  (see Appendix 1) with
respect to the scalar product (\ref{eq:E2.14}):
\begin{eqnarray}  
V^{(1)}_{{\vec k},\nu}(x)={e^{-\pi\nu/2}\over 2^{5/2}\pi k_{\bot}} 
\left( \begin{array}{c} 
                         k_y \\ 
                        -k_x \\ 
                         0 
                             \end{array} \right)
H^{(2)}_{-i\nu} (k_{\bot}\tau) e^{i\nu\eta +i{\vec k}{\vec r}};\;\;\; 
{\rm and}\;\;\;
V^{(2)}_{{\vec k},\nu}(x)={e^{-\pi\nu/2}\over 2^{5/2}\pi k_{\bot}} 
\left( \begin{array}{c} 
                 \nu k_x R^{(2)}_{-1,-i\nu}(k_{\bot}\tau) \\ 
                 \nu k_y R^{(2)}_{-1,-i\nu}(k_{\bot}\tau) \\ 
                 - R^{(2)}_{1,-i\nu}(k_{\bot}\tau) 
                                            \end{array} \right)
                e^{i\nu\eta +i{\vec k}{\vec r}}~.
\label{eq:E2.15}\end{eqnarray}                                  
The mode $V^{(2)}$ is constructed from the functions 
$R^{(j)}_{\mu,-i\nu}(k_{\bot}\tau)=R^{(j)}_{\mu,-i\nu}(k_{\bot}\tau|s)$
corresponding to the boundary condition of vanishing gauge field at
$\tau=0$. This guarantee continuous behavior of the field at $\tau=0$.
Indeed, at $\tau\rightarrow 0$,~ the normal and the tangent directions
become degenerate. As long as $A^\tau=0$ is the gauge condition, the
continuity requires that  $A^\eta\rightarrow 0$ at $\tau\rightarrow 0$.

In order to simplify some  subsequent calculations it is useful to 
write down explicitly the physical components of electric and magnetic 
fields of these modes, 
$ {\cal E}^m=\sqrt{-{\rm g}}{\rm g}^{mn}\stackrel{\bullet}{A}_n$ 
and  $ {\cal B}^m=-2^{-1}e^{mln}F_{ln}$; 
\begin{eqnarray}  
{\cal E}^{(1)m}_{{\vec k},\nu}(x)=i{\cal B}^{(2)m}_{{\vec k},\nu}(x)
={e^{-\pi\nu/2}\over 2^{5/2}\pi k_{\bot}} 
\left( \begin{array}{c}  k_y \\ 
                        -k_x \\ 
                         0   \end{array} \right)
\stackrel{\bullet}{H}^{(2)}_{-i\nu} (k_{\bot}\tau) 
e^{i\nu\eta +i{\vec k}{\vec r}}~~, 
\nonumber\\
{\cal E}^{(2)m}_{{\vec k},\nu}(x)=i{\cal B}^{(1)m}_{{\vec k},\nu}(x)
={e^{-\pi\nu/2}\over 2^{5/2}\pi k_{\bot}} 
\left( \begin{array}{c}  \nu k_x  \\ 
                         \nu k_y  \\ 
                         - k_{\bot}^{2} \end{array} \right)
H^{(2)}_{-i\nu} (k_{\bot}\tau) e^{i\nu\eta +i{\vec k}{\vec r}}~.
\label{eq:E2.15A}\end{eqnarray}    
Therefore, the mode $V^{(2)}$ can be obtained from the mode $V^{(1)}$
by simple interchange of its electric and magnetic fields. Using the
standard wave-guide terminology, one may call mode $V^{(1)}$  as the
``transverse electric mode'' and the mode $V^{(2)}$ -- as the ``transverse
magnetic mode''.

An equivalent full set of the transverse modes carries  instead of the 
boost $\nu$, the quantum number $\theta$, rapidity: 
$k_0= k_{\bot}\cosh\theta,~~k_3= k_{\bot} \sinh\theta$.
These functions can be obtained by means of the Fourier transform,
\begin{eqnarray}  
v^{(\lambda)}_{{\vec k},\theta}(x)
= \int_{-\infty}^{+\infty}  {d\nu \over (2\pi)^{1/2}i} e^{i\nu\theta}
V^{(\lambda)}_{{\vec k},\nu}(x)~~,
\label{eq:E2.15a} \end{eqnarray}        
and have the following form,
\begin{eqnarray}  
v^{(1)}_{{\vec k},\theta}(x)={1\over 4\pi^{3/2} k_{\bot}} 
\left( \begin{array}{c} 
                         k_y \\ 
                        -k_x \\ 
                         0 
                             \end{array} \right)
e^{-ik_{\bot}\tau\cosh (\theta-\eta) +i{\vec k}{\vec r}}; \;\;\;
v^{(2)}_{{\vec k},\theta}(x)={1\over 4\pi^{3/2} k_{\bot}}  
\left( \begin{array}{c} 
                 k_x f_1 \\ 
                 k_y f_1 \\ 
                  - f_2 
                 \end{array} \right) e^{i{\vec k}{\vec r}}~~,
\label{eq:E2.15b}\end{eqnarray}  
where
\begin{eqnarray}  
f_1(\tau ,\eta)=  k_{\bot} \sinh (\theta-\eta) \int_{0}^{\tau} 
e^{-ik_{\bot}\tau' \cosh (\theta-\eta)}d\tau' =i \tanh (\theta-\eta)
(e^{-ik_{\bot}\tau \cosh (\theta-\eta)} - 1)~~, \nonumber
\end{eqnarray}                                             
\begin{eqnarray}  
f_2(\tau ,\eta)= k_{\bot}^2\int_{0}^{\tau}
e^{-ik_{\bot}\tau' \cosh (\theta-\eta)}\tau' d\tau'=
{e^{-ik_{\bot}\tau \cosh (\theta-\eta)} -1 \over
\cosh^{2} (\theta-\eta)} +
i k_{\bot}\tau {e^{-ik_{\bot}\tau\cosh (\theta-\eta)} 
\over  \cosh (\theta-\eta)}~~,
\label{eq:E2.15c} 
\end{eqnarray}        
The norm of the Coulomb mode $V^{(3)}$, as defined by Eq.~(\ref{eq:E2.14}),
equals zero,
and it is orthogonal to $V^{(1)}$ and $V^{(2)}$. Though this solution 
obeys equations of motion without the current, it does not obey the Gauss
law without a charge. Therefore, it should be discarded in decomposition
of the radiation field. However, it should have been kept if we considered
the radiation field in the presence of the static source with the
$\tau$-independent density  
$\rho({\vec k},\nu)=\tau j^{\tau}_{{\vec k}\nu}(\tau)=const(\tau) $. In this
case its definition can be completed using the Gauss law:
\begin{eqnarray} 
V^{(3)}_{{\vec k},\nu}(x)={ \rho({\vec k},\nu) \over (2\pi)^3 i k_{\bot}^2}
 \left( \begin{array}{c} 
                 k_r  Q_{-1,i\nu}(k_{\bot}\tau) \\ 
                 \nu~Q_{1,i\nu}(k_{\bot}\tau)
                                            \end{array} \right) 
                   e^{i\nu\eta +i{\vec k}{\vec r}}~~.     
\label{eq:E2.16}\end{eqnarray}                     
The coordinate form of this solution is noteworthy. The physical
components, ~${\cal E}^m=\sqrt{-{\rm g}}{\rm g}^{ml}{\stackrel{\bullet} A}_l$~, of
the electric field of the ``$\tau$--static'' source can be written down in
the integral form,
\begin{eqnarray}   
{\cal E}_{(stat)}^{i}(\tau,{\vec r}_1,\eta_1)= \int d{\vec r}_2 d\eta_2 
K_i(\tau;{\vec r}_1-{\vec r}_2,\eta_1 -\eta_2) \rho({\vec r}_2,\eta_2)~,
\label{eq:E2.17}\end{eqnarray}   
with the kernel
\begin{eqnarray}   
K_i(\tau;{\vec r},\eta)= \int {d\nu d^2{\vec k}\over (2\pi)^3}
{e^{i\nu\eta +i{\vec k}{\vec r}}\over i k_{\bot}^2} 
\left( \begin{array}{c} 
                 k_r   s_{1,i\nu}(k_{\bot}\tau) \\ 
                 \nu k_{\bot}^2  s_{-1,i\nu}(k_{\bot}\tau)
                                            \end{array} \right) 
              = -{\theta(\tau -r_{\bot})\over 4\pi}
\left( \begin{array}{c} 
                   \tau\cosh\eta(\partial / \partial x^r) \\ 
                    \partial / \partial(\tau\sinh\eta)
                           \end{array} \right) {1\over R_{12}}~~,
\label{eq:E2.18}\end{eqnarray}
where
\begin{eqnarray}   
R_{12}=(r_{\bot}^2+\tau^2\sinh^2\eta)^{1/2}=
[({\vec r}_1-{\vec r}_2)^2-\tau^2\sinh^2(\eta_1 -\eta_2)]^{1/2}~,   
\label{eq:E2.18a}\end{eqnarray} 
is the distance between the points ${\vec r}_1,\eta_1$ and 
${\vec r}_2,\eta_2$  in the internal geometry of the surface $\tau=const$.
The technical details of derivation of the last expression will be adduced
in Sec.~\ref{sec:SN5}. Eq.~(\ref{eq:E2.18}) is an analogue of the Coulomb
law of electrostatics, except that now the source has a  density which is
static with respect to the Hamiltonian time $\tau$. In fact, the source is
static if it expands in such a way that $\tau j^\tau(\tau, \eta, {\vec
r})$ does not depend on $\tau$.   These expressions  will be helpful in 
recognizing the origin of  various terms in the full propagator which
is calculated below.

\subsection{Propagator in the gauge $A^\tau=0$}
\label{subsec:SB23}  

Calculation of the propagator in the gauge $A^\tau=0$ associated with the
system of the curved surfaces $\tau=const$  meets several problems. Three
methods are commonly used in the field theory. One of them strongly
appeals to the Fourier analysis in the plane Minkowsky space which is not
applicable now because the metric itself is coordinate-dependent. The
second method uses the path-integral formulation which is also ineffective
because of the explicit coordinate dependence of the gauge-fixing term in
the Lagrangian. One could also try to study the spectrum of the matrix 
differential operator, to find its eigen-functions and to use the standard
expression for the  resolvent. However, the  extension of the system for
the non-zero eigenvalues leads to the unwieldy equations.  On the other
hand, the Green function of the perturbation theory must coincide with the
one which solves the problem of the gauge field interaction with the
classical ``external'' current.  For this reason, we shall compute the
Green function in a most straightforward way: we shall look for the
partial solution of the inhomogeneous system using the method of
``variation of parameters''. 

Let us start derivation of the propagator in the gauge $A^\tau=0$ with
obtaining the separate differential equations for the $\eta$-component of
the magnetic field, $\Psi= \partial_{y}A_x -\partial_{x} A_y $; the
transverse divergence of the electric field,   $\varphi= \tau
(\partial_{x}{\stackrel{\bullet} A}_x +\partial_{y}{\stackrel{\bullet} A}_y) $; and the
$\eta$-component of the electric field, $a={\stackrel{\bullet} A}_\eta /\tau$~. In terms
of the Fourier components with respect to the spacial coordinates, these
equations read as
\begin{eqnarray}  
[\partial_{\tau}^{2}+{1\over\tau}\partial_{\tau}+ {\nu^2\over\tau^2}+
k_{\bot}^2] \Psi_{{\vec k},\nu}(\tau)= -j^\psi({\vec k},\nu,\tau)~~, 
\label{eq:E2.19}\end{eqnarray}   
\begin{eqnarray}
[\partial_{\tau}\tau\partial_{\tau}+
{\nu^2 \over \tau}] \varphi ({\vec k},\nu,\tau) -i\tau \nu k_{\bot}^{2}
a ({\vec k},\nu,\tau)
=-\partial_\tau [\tau^2 j^\varphi({\vec k},\nu,\tau) ] ~~,
\label{eq:E2.20}\end{eqnarray}             
\begin{eqnarray}
[\partial_{\tau}\tau\partial_{\tau}+\tau k_{\bot}^{2}] a({\vec k},\nu,\tau)
-{i\nu \over \tau} \varphi({\vec k},\nu,\tau) 
=-\partial_\tau [\tau^2 j^\eta ({\vec k},\nu,\tau)] ~~,              
\label{eq:E2.21}\end{eqnarray}   
where $j^\psi=\partial_{y}j_x -\partial_{x} j_y $,~
 $j^\varphi=\partial_{x}j_x +\partial_{y} j_y $.
Using  the constraint conservation, which may be explicitly integrated to
\begin{eqnarray}
 \varphi({\vec k},\nu,\tau) )+i\nu a({\vec k},\nu,\tau)- 
\tau j^\tau ({\vec k},\nu,\tau) = -\rho_0({\vec k},\nu)=const(\tau) ~~,  
\label{eq:E2.22}\end{eqnarray}   
one easily obtains two independent equations for $\varphi({\vec k},\nu,\tau)$
and $a({\vec k},\nu,\tau)$: 
\begin{eqnarray}
[\partial_{\tau}^{2}+{1\over \tau}\partial_{\tau}+
{\nu^2 \over \tau^2} + k_{\bot}^{2}]\varphi ({\vec k},\nu,\tau)
=k_{\bot}^2[\rho({\vec k},\nu,\tau)-\rho_0({\vec k},\nu)]-
{1\over \tau}\partial_\tau (\tau^2 j^\varphi({\vec k},\nu,\tau) )   
          \equiv f^\varphi ~~,
\label{eq:E2.23}\end{eqnarray}             
\begin{eqnarray}
[\partial_{\tau}^{2}+{1\over \tau}\partial_{\tau}+
{\nu^2 \over \tau^2} + k_{\bot}^{2}] a({\vec k},\nu,\tau)
={-i\nu\over \tau^2}[\rho({\vec k},\nu,\tau)-\rho_0({\vec k},\nu)]-
{1\over \tau}\partial_\tau (\tau^2 j^\eta ({\vec k},\nu,\tau) ) 
\equiv f^\eta ~~,              
\label{eq:E2.24}\end{eqnarray} 
The constant of integration $\rho_0({\vec k},\nu)$  has meaning of the
arbitrary static charge density and it should be retained until the Gauss law
is explicitly imposed on the solution. In what follows, we shall not
write it down explicitly assuming that it is included in the true
charge density $\rho({\vec k},\nu,\tau)$. Equations~(\ref{eq:E2.19}),
(\ref{eq:E2.23}) and (\ref{eq:E2.24}) can be solved by the method of
``variation of parameters'':     
\begin{eqnarray}
{\cal F}(\tau)= {\pi i \over 4} \int_{0}^{\tau}\tau_2d\tau_2
{\cal H}(\tau,\tau_2) f(\tau_2)~~,
\label{eq:E2.25}\end{eqnarray}     
where ${\cal F}$ stands for anyone of the unknown functions in these 
equations, and $f$ for the corresponding right hand side.  The kernel
\begin{eqnarray}
{\cal H}(\tau,\tau_2)= 
H^{(1)}_{i\nu}(k_{\bot}\tau) H^{(2)}_{i\nu}(k_{\bot}\tau_2)
-H^{(2)}_{i\nu}(k_{\bot}\tau) H^{(1)}_{i\nu}(k_{\bot}\tau_2)~~.\nonumber
\end{eqnarray}
is a usual bilinear form built from the linearly independent solutions of the
homogeneous equation. [ The Wronskian of these solutions is exactly $4/
i\pi \tau_2$. ] Taking  ${\cal F}=\Psi$,  we obtain the first equation for
the components $A_x({\vec k},\nu,\tau)$ and $A_y({\vec k},\nu,\tau)$ of
the vector  potential:
\begin{eqnarray}
\Psi({\vec k},\nu,\tau_1)\equiv
i[-k_y A_x + k_x A_y]= {i\pi\over 4} \int_{0}^{\tau}\tau_2d\tau_2
{\cal H}(\tau_1,\tau_2)~ i~[-k_y j^x(\tau_2) + k_x j^y(\tau_2)]~~.
\label{eq:E2.26}\end{eqnarray}                                  
In order to find the second equation for the $x$- and $y$--components 
and the equation for $A_\eta({\vec k},\nu,\tau)$ we must integrate twice:
\begin{eqnarray}
\Phi({\vec k},\nu,\tau_1)\equiv
i[k_x A_x + k_y A_y]= {i\pi\over 4} 
\int_{0}^{\tau_1}{d \tau' \over \tau'} \int_{0}^{\tau'} {\cal H}(\tau',\tau_2)
\tau_2 d\tau_2  ~\big[ -k_{\bot}^2 ~\rho({\vec k},\nu,\tau_2)+
{1\over \tau_2}\partial_{\tau_2} \big(\tau_{2}^{2}
j^\varphi({\vec k},\nu,\tau_2) \big)\big] ~~,
\label{eq:E2.27}\end{eqnarray}                                  
\begin{eqnarray}
A_\eta({\vec k},\nu,\tau_1) = {i\pi\over 4} 
\int_{0}^{\tau_1} \tau' d \tau' \int_{0}^{\tau'} {\cal H}(\tau',\tau_2)
\tau_2 d\tau_2 ~ \big[{i\nu\over \tau_{2}^{2}} ~\rho({\vec k},\nu,\tau_2)+
{1\over \tau_2}\partial_{\tau_2} \big(\tau_{2}^{2}
j^\eta({\vec k},\nu,\tau_2) \big)\big] ~~.
\label{eq:E2.28}\end{eqnarray}              
Here, the integration over $\tau_2$ recovers the electric fields at the
moment $\tau'$, whilst the integration over $\tau'$ gives the vector
potential at the moment $\tau_1$. It is convenient to start with the latter
one which has the limits  $\tau_2 <\tau'<\tau_1$. Let us consider the main
line of calculations in detail,
using the $\eta$-component as an example.
The first integration follows the formula (\ref{eq:A2.1}),
\begin{eqnarray}   
k_{\bot}^{\mu +1}\int_{\tau_2}^{\tau_1} (\tau')^\mu 
H^{(j)}_{i\nu}(k_{\bot}\tau') d \tau' = 
R^{(j)}_{\mu,i\nu}(k_{\bot}\tau_1)-R^{(j)}_{\mu,i\nu}(k_{\bot}\tau_2)~~,
\label{eq:E2.29}\end{eqnarray}    
and  the terms emerging from the lower limit $\tau_2$ can be conveniently 
transformed according to the relation (see Appendix~2),
\begin{eqnarray}   
R^{(1)}_{\mu,i\nu}(k_{\bot}\tau_2)H^{(2)}_{i\nu}(k_{\bot}\tau_2) -
R^{(2)}_{\mu,i\nu}(k_{\bot}\tau_2)H^{(1)}_{i\nu}(k_{\bot}\tau_2)     
={4\over i\pi}s_{\mu,i\nu}(k_{\bot}\tau_2)~~
\label{eq:E2.30}\end{eqnarray}      
As a result, one obtains, {\em e.g.}, the following formula for 
 $A_\eta({\vec k},\nu,\tau_1)$:
\begin{eqnarray}
A_\eta({\vec k},\nu,\tau_1) = {i\pi\over 4 k_{\bot}^2}
\int_{0}^{\tau_1}\tau_2 d\tau_2
[R^{(1)}_{1,i\nu}(k_{\bot}\tau_1)H^{(2)}_{i\nu}(k_{\bot}\tau_2) -
R^{(2)}_{1,i\nu}(k_{\bot}\tau_1)H^{(1)}_{i\nu}(k_{\bot}\tau_2) 
- {4\over i\pi}s_{1,i\nu}(k_{\bot}\tau_2)] \nonumber \\  
\times \big[{i\nu\over \tau_{2}^{2}} \rho({\vec k},\nu,\tau_2)+
{1\over \tau_2}\partial_{\tau_2} \big(\tau_{2}^{2}
j^\eta({\vec k},\nu,\tau_2) \big) \big]~~.
\label{eq:E2.31}\end{eqnarray}          
In order to eliminate the charge density $\rho$ from the integrand and to
separate the transverse and the longitudinal parts of the propagator, all
the terms of this formula should be integrated by parts with explicit
account for the charge conservation which reads as
\begin{eqnarray}
i\tau [k_x j^x({\vec k},\nu,\tau) + k_x j^y({\vec k},\nu,\tau)
+\nu j^\eta({\vec k},\nu,\tau)]+\partial_\tau \rho({\vec k},\nu,\tau)=0~~.
\label{eq:E2.32}\end{eqnarray}  
We have in sequence:
\begin{eqnarray}   
i\nu\int_{0}^{\tau_1} {d\tau_2\over\tau_2} \rho(\tau_2)
H^{(j)}_{i\nu}(k_{\bot}\tau_2) = i\nu\int_{0}^{\tau_1} 
{d R^{(j)}_{-1,i\nu}(k_{\bot}\tau_2) \over d\tau_2} 
\rho(\tau_2)d\tau_2 \nonumber\\ 
=i\nu R^{(j)}_{-1,i\nu}(k_{\bot}\tau_1) \rho(\tau_1)-
\nu \int_{0}^{\tau_1} \tau_2 d\tau_2 R^{(j)}_{-1,i\nu}(k_{\bot}\tau_2)      
[k_x j^x(\tau_2) + k_y j^y(\tau_2)+\nu j^\eta(\tau_2)]~~,
\label{eq:E2.33}\end{eqnarray}        
\begin{eqnarray}   
i\nu\int_{0}^{\tau_1} d\tau_2 
H^{(j)}_{i\nu}(k_{\bot}\tau_2)\partial_{\tau_2} (\tau_{2}^{2}
j^\eta(\tau_2) ) 
= \tau_{1}^{2} j^\eta(\tau_1)H^{(j)}_{i\nu}(k_{\bot}\tau_1)  
+ \int_{0}^{\tau_1} \tau_2 d\tau_2 [R^{(j)}_{1,i\nu}(k_{\bot}\tau_2)      
 +\nu^2 R^{(j)}_{-1,i\nu}(k_{\bot}\tau_2)]j^\eta(\tau_2).
\label{eq:E2.34}\end{eqnarray}        
In a similar way we have,
\begin{eqnarray}   
i\nu\int_{0}^{\tau_1} {d\tau_2\over\tau_2} \rho(\tau_2)
s_{1,i\nu}(k_{\bot}\tau_2) = i\nu\int_{0}^{\tau_1}
{dQ_{-1,i\nu}(k_{\bot}\tau_2)\over d\tau_2} \rho(\tau_2)d\tau_2 \nonumber\\ 
=i\nu Q_{-1,i\nu}(k_{\bot}\tau_1) \rho(\tau_1)-
\nu \int_{0}^{\tau_1} \!\! \tau_2 d\tau_2 Q_{-1,i\nu}(k_{\bot}\tau_2)      
[k_x j^x(\tau_2) + k_y j^y(\tau_2)+\nu j^\eta(\tau_2)]~~,
\label{eq:E2.35}\end{eqnarray}       
\begin{eqnarray}   
\int_{0}^{\tau_1} d\tau_2 s_{1,i\nu}(k_{\bot}\tau_2)
\partial_{\tau_2} (\tau_{2}^{2}j^\eta(\tau_2) ) 
= \tau_{1}^{2}
j^\eta(\tau_1)s_{1,i\nu}(k_{\bot}\tau_1)  
+ \nu^2 \int_{0}^{\tau_1} \tau_2 d\tau_2 [Q_{-1,i\nu}(k_{\bot}\tau_2)      
-Q_{1,i\nu}(k_{\bot}\tau_2)]j^\eta(\tau_2)~~.
\label{eq:E2.36}\end{eqnarray}  
Assembling these pieces together and repeating the same calculations
for the function $\Phi$ one obtains three different terms which contribute
to the field $A$ produced by the current $j$:~~
$A=A^{(tr)}+A^{(L)}+A^{(inst)}$.

The transverse field $A^{(tr)}$ is defined by the integral terms in 
the R.H.S. of Eqs.~(\ref{eq:E2.33}) and (\ref{eq:E2.34}). It can be
conveniently written down in the following form:
\begin{eqnarray}
A^{(tr)}_{l}(x_1)=\int d^4 x_2 \theta(\tau_1-\tau_2)
\Delta^{(tr)}_{lm}(x_1,x_2) j^m(x_2)~~.
\label{eq:E2.37}\end{eqnarray}     
where                             
\begin{eqnarray}
\Delta^{(tr)}_{lm}(x,y)= -i\int_{-\infty}^{\infty} 
d\nu\int d^2{\vec k}\sum_{\lambda=1,2}
[V^{(\lambda)}_{\nu {\vec k};l}(x) V^{(\lambda)\ast }_{\nu {\vec k};m}(y)-
V^{(\lambda)\ast}_{\nu {\vec k};l}(x) V^{(\lambda) }_{\nu {\vec k};m}(y)]~~,
\label{eq:E2.38}\end{eqnarray}   
can be easily recognized as the Riemann function of the original
homogeneous hyperbolic system. The Riemann function solves the boundary
value problem for the evolution of the  free radiation field.  It is
obtained immediately in the form of the bilinear expansion over the full
set of the solutions  (\ref{eq:E2.15}) of the homogeneous system. In fact,
this is a sole evidence that  $\Delta^{(tr)}$ may be associated with the 
transverse part of the propagator. Then the rest part is the 
propagator (response function) for the longitudinal field.

The dynamical  longitudinal field $A^{(L)}$ originates from the integral 
terms in  the R.H.S. of Eqs.~(\ref{eq:E2.35}) and (\ref{eq:E2.36}): 
\begin{eqnarray}
A^{(L)}_{l}(\tau_1,\eta_1,{\vec r}_1)=\int_{0}^{\tau_1} \tau_2 d\tau_2 
\int  d\eta_2 d^2{\vec r}_2
\Delta^{(L)}_{lm}(\tau_2;\eta_1-\eta_2,{\vec r}_1-{\vec r}_2) 
j^m(\tau_2,\eta_2,{\vec r}_2)~~.     
\label{eq:E2.39}\end{eqnarray}         
and the kernel of this representation,        
\begin{eqnarray}
\Delta^{(L)}_{lm}(\tau_2;\eta_1-\eta_2,{\vec r}_1-{\vec r}_2) 
= \int {d\nu d^2{\vec k} \over (2\pi)^3 k_{\bot}^2}
 \left[ \begin{array}{c} 
                 k_r  \\ 
                 \nu  \end{array} \right]_l 
\left[ \begin{array}{c} 
                 k_s  Q_{-1,i\nu}(k_{\bot}\tau_2) \\ 
                 \nu Q_{1,i\nu}(k_{\bot}\tau_2)
                                            \end{array} \right]_m 
              e^{i\nu(\eta_1-\eta_2) +i{\vec k}({\vec r}_1-{\vec r}_2)}~~,     
\label{eq:E2.40}\end{eqnarray}           
does not allow for the bilinear expansion with two temporal arguments,
and, as we shall see in a while, the retarded character of the integration
in Eq.~(\ref{eq:E2.39}) is not sensitive to the light cone
boundaries. In fact, the
electric field $E_{l}^{(L)}={\stackrel{\bullet} A}_{l}^{(L)}$ is simultaneous with the 
current $j^m$ .

The last one, instantaneous, part of the solution comes from the 
boundary terms in Eqs.~(\ref{eq:E2.33})--(\ref{eq:E2.36}) which were
generated via  integration by parts.  It depends on a single time
variable $\tau_1$. Using two functional relations, (\ref{eq:E2.30}) and
\begin{eqnarray}   
R^{(1)}_{1,i\nu}(x)R^{(2)}_{-1,i\nu}(x)   -
R^{(2)}_{1,i\nu}(x)R^{(1)}_{-1,i\nu}(x)       
=-{4\over i\pi}{x\over\nu^2}{d s_{1,i\nu}(x)\over d x}=
-{4\over i\pi}[Q_{1,i\nu}(x)-Q_{-1,i\nu}(x)] ~~,
\label{eq:E2.41}\end{eqnarray}      
( see Appendix~2 ) its Fourier transform can be presented in the form 
\begin{eqnarray} 
A^{(inst)}_{l}({\vec k},\nu;\tau_1)=
{ \rho({\vec k},\nu,\tau_1) \over (2\pi)^3 i k_{\bot}^2}
 \left[ \begin{array}{c} 
                 k_r  Q_{-1,i\nu}(k_{\bot}\tau_1) \\ 
                 \nu Q_{1,i\nu}(k_{\bot}\tau_1)
                                            \end{array} \right]_l~~,     
\label{eq:E2.42}\end{eqnarray}                     
which leads to the Poisson-type integral,
\begin{eqnarray} 
A^{(inst)}_{m}(\tau_1,\eta_1,{\vec r}_1) = \int d{\vec r}_2 d\eta_2 
{\cal K}_m(\tau_1;{\vec r}_1-{\vec r}_2,\eta_1 -\eta_2) 
\rho(\tau_1,{\vec r}_2,\eta_2)~,
\label{eq:E2.43}\end{eqnarray}    
with the {\em instantaneous} kernel,
\begin{eqnarray}   
{\cal K}_m(\tau;{\vec r},\eta)= \int {d\nu d{\vec k}\over (2\pi)^3}
{e^{i\nu\eta +i{\vec k}{\vec r}}\over i k_{\bot}^2}
\left[ \begin{array}{c} 
                 k_r   Q_{-1,i\nu}(k_{\bot}\tau) \\ 
                 \nu  Q_{1,i\nu}(k_{\bot}\tau)
                                            \end{array} \right]_m~~.
\label{eq:E2.44}\end{eqnarray}                                          
Therefore, this term represents instantaneous distribution of the
potential at the moment $\tau_1$, corresponding to the charge density
taken at the same moment. Recalling that the charge density   $\rho({\vec
k},\nu,\tau_1) $ in Eq.~(\ref{eq:E2.42}) still includes the arbitrary
constant $\rho_0({\vec k},\nu)$, we see that  imposing the constraint
indeed affects only the potential of static charge distribution and puts
it in agreement with the Gauss law. If $\rho_0$ is set to zero, then the
gauge is completely fixed and all calculations with this propagator will
be gauge invariant. 

For the practical calculation, it is easier to work with the components 
$j^n$ of the current rather than to keep the charge density $\rho$ in its 
original form. Otherwise, {\em e.g.}, the expression for the vertex
function will be unwieldy.
Replacement follows the prescription: 
\begin{eqnarray}
\rho(\tau_1,\eta_2,{\vec r}_2)= \int_{0}^{\tau_1}d\tau_2 {\partial\rho
\over\partial\tau_2} = -i \int_{0}^{\tau_1}\tau_2 d\tau_2
[k_s j^s(\tau_2,\eta_2,{\vec r}_2)+
\nu j^\eta(\tau_2,\eta_2,{\vec r}_2)]~~, \nonumber   
\end{eqnarray} 
and restores an extraneous ``initial'' configuration of  the static charge
which has been previously removed from Eq.~(\ref{eq:E2.42}). Being
$\tau$-dependent, its vector potential cannot be eliminated by the
residual  gauge transformation. However, this artificial contribution
corresponds to the easily recognizable static pattern in the longitudinal
part of the propagator and is under the full control. Keeping this
fragment in mind, we arrive at the standard form of the representation,
\begin{eqnarray}
A^{(inst)}_{l}(\tau_1,\eta_1,{\vec r}_1)=\int_{0}^{\tau_1} \tau_2 d\tau_2 
\int  d\eta_2 d^2{\vec r}_2
\Delta^{(inst)}_{lm}(\tau_1;\eta_1-\eta_2,{\vec r}_1-{\vec r}_2) 
j^m(\tau_2,\eta_2,{\vec r}_2)~~.     
\label{eq:E2.43a}\end{eqnarray}    
with the kernel given by the formula,
\begin{eqnarray}
\Delta^{(inst)}_{lm}(\tau_1;\eta_1-\eta_2,{\vec r}_1-{\vec r}_2 )
= - \int {d\nu d^2{\vec k} \over (2\pi)^3 k_{\bot}^2}
\left[ \begin{array}{c} 
                 k_r  Q_{-1,i\nu}(k_{\bot}\tau_1) \\ 
                 \nu Q_{1,i\nu}(k_{\bot}\tau_1)
                                            \end{array} \right]_l 
 \left[ \begin{array}{c} 
                 k_s  \\ 
                 \nu  \end{array} \right]_m    
               e^{i\nu(\eta_1-\eta_2) +i{\vec k}({\vec r}_1-{\vec r}_2)}~~.     
\label{eq:E2.44a}\end{eqnarray}              

Eqs. (\ref{eq:E2.37})--(\ref{eq:E2.40}) and (\ref{eq:E2.43a}), 
(\ref{eq:E2.44a}) present the propagator in a split form. Different
constituents of this form are preliminary identified as transverse, 
longitudinal and instantaneous parts of the propagator. It would be useful
to learn if the same kind of splitting is possible for the current itself.
An affirmative answer (as in the cases of the Coulomb and radiation
gauges) would be  helpful for the design of the perturbation theory.  To
answer the question, one should substitute the different pieces of the
solution  into the left hand side of the original system of differential
equations. This leads to the following expressions for the Fourier
components of the three currents:
\begin{eqnarray}   
\tau j^{m}_{(tr)}({\vec k},\nu;\tau)=\tau j^{m}({\vec k},\nu;\tau)+
 {1\over ik_{\bot}^{2}}     \left[ \begin{array}{c} 
    k_r   s_{1,i\nu}(k_{\bot}\tau) \\ 
    \nu k_{\bot}^{2}  s_{-1,i\nu}(k_{\bot}\tau)  
    \end{array} \right]^m {\partial\rho\over\partial\tau} -
{\nu\over k_{\bot}^{2}} {\partial\over\partial\tau}
               \bigg( {\stackrel{\bullet} s}_{-1,i\nu}(k_{\bot}\tau)
\left[ \begin{array}{c} 
                 k_r  \tau^3 j^\eta  \\ 
                 -\tau (k_xj^x+k_yj^y)
                             \end{array} \right]^m \bigg)~~,
\label{eq:E2.45}\end{eqnarray} 
\begin{eqnarray}   
\tau j^{m}_{(L)}({\vec k},\nu;\tau)={1\over k_{\bot}^{2}} 
 {\partial\over\partial\tau} \bigg(
\left[ \begin{array}{c} 
                 k_r\tau^2 \\ \nu \end{array} \right]^m 
[Q_{-1,i\nu}(k_{\bot}\tau)(k_xj^x+k_yj^y) 
+Q_{1,i\nu}(k_{\bot}\tau) j^\eta ]\bigg)~~,
\label{eq:E2.45a}\end{eqnarray} 
\begin{eqnarray}   
\tau j^{m}_{(inst)}({\vec k},\nu;\tau)= {-1\over ik_{\bot}^{2}}
        {\partial\over\partial\tau} 
    \bigg( \left[ \begin{array}{c} 
                 k_r   \tau Q_{-1,i\nu}(k_{\bot}\tau) \\ 
                 \nu  \tau^{-1} Q_{1,i\nu}(k_{\bot}\tau)
                  \end{array} \right]^m
{\partial\rho\over\partial\tau}  \bigg) -{1\over ik_{\bot}^{2}}\ 
\left[ \begin{array}{c} 
                 k_r   s_{1,i\nu}(k_{\bot}\tau) \\ 
                 \nu k_{\bot}^{2}  s_{-1,i\nu}(k_{\bot}\tau)
                                            \end{array} \right]^m 
                       {\partial\rho\over\partial\tau}~~, 
\label{eq:E2.46}\end{eqnarray}           
Providing the current is conserved, these three currents, added together,
give the full current from the right hand side of the system. Therefore,
the solution is correct. However, none of these three currents  carries
any signature of being longitudinal or transversal in a usual sense. None
of them  has zero divergence, since the operator of the divergence does
not commute with the differential operator of the system. No desired
simplification is possible in our case.

In fact, even the above splitting of the potential has no real physical
meaning. To see it explicitly, let us find the divergence of the electric
field, ${\rm div}{\bf E}= \partial_m{\cal E}^m $ [again, for brevity, in the
Fourier representation]:
\begin{eqnarray}  
{\rm div}{\bf E}^{(tr)}({\vec k},\nu;\tau) 
=i[Q_{-1,i\nu}(k_{\bot}\tau)-Q_{1,i\nu}(k_{\bot}\tau)]
\big(\nu\tau^2 j^\eta - {\nu^2\over k_{\bot}^{2}}(k_xj^x+k_yj^y)\big)~~, 
\label{eq:E2.46A}\end{eqnarray}  
\begin{eqnarray}  
{\rm div}{\bf E}^{(L)}({\vec k},\nu;\tau)
=i \big( \tau^2+{\nu^2\over k_{\bot}^{2}}\big)
[(k_xj^x+k_yj^y) Q_{-1,i\nu}(k_{\bot}\tau)
+\nu j^\eta Q_{1,i\nu}(k_{\bot}\tau)]~~,
\label{eq:E2.46B}\end{eqnarray}  
\begin{eqnarray}  
{\rm div}{\bf E}^{(inst)}({\vec k},\nu;\tau)=
\rho({\vec k},\nu;\tau) -i \big( \tau^2 Q_{-1,i\nu}(k_{\bot}\tau) 
-{\nu^2\over k_{\bot}^{2}} Q_{1,i\nu}(k_{\bot}\tau)\big)
[(k_xj^x+k_yj^y) +\nu j^\eta]~~.
\label{eq:E2.46C}\end{eqnarray}  
Only the divergence of the true retarded component of the field
${\bf E}^{(tr)}$ turns out to be zero. The term which prevents   
${\rm div}{\bf E}^{(tr)}$ from being zero is due to non-symmetry of  the
propagator,  $\Delta^{\eta r}\neq\Delta^{r\eta}$. It appears when the
$\theta$-function in Eq.~(\ref{eq:E2.37}) is  differentiated with respect to
Hamiltonian  time $\tau$. This term is vital for obtaining  the expression
that obeys Gauss law constraint, ${\rm div}{\bf E}({\vec k},\nu;\tau)=
\rho({\vec k},\nu;\tau)$.  [ We remind that $\rho({\vec k},\nu;\tau)$  
still includes an arbitrary  constant $\rho_o({\vec k},\nu)$.] 

The known examples, when the transverse and  the longitudinal fields are
separated at the level of equations of motion, are  related to the narrow
class of homogeneous gauges. 
Impossibility to perform a universal
separation of the transverse and longitudinal fields thus appears to be a
rule rather than exception. It reflects a general principle: the radiation
field created at some time interval has the preceeding and the subsequent
configurations of the longitudinal field as the boundary condition.
Dynamics of the longitudinal field falls out of any scattering problem in its
$S$-matrix formulation. However, this dynamics is, in fact, a subject of
the QCD evolution in deep inelastic scattering

\subsection{Initial data problem in the gauge $A^{\tau}=0$} 
\label{subsec:SB25}

We obtained the expression for the (retarded) propagator as the response 
function between the ``external'' current and the potential of the gauge 
field. We must verify that the same propagator solves the Cauchy
problem for the gauge field. This can be easily done by presenting
the initial data at the surface $\tau=\tau_0$ in the form of the 
source density at the hypersurface $\tau=\tau_0$,
\begin{eqnarray} 
\sqrt{-{\rm g}}J^n(\tau_2)=\sqrt{-{\rm g}(\tau_0)}{\rm g}^{nm}(\tau_0)
[\delta'(\tau_2-\tau_0){\bar A}_m({\vec r},\eta)+
\delta(\tau_2-\tau_0){\bar A'}_m({\vec r},\eta) ]~,
\label{eq:E2.47}\end{eqnarray}
where ${\bar A}_m({\vec r},\eta)$ and ${\bar A'}_m({\vec r},\eta)$ are
the initial data for the potential and its normal derivative on the
hypersurface $\tau = \tau_0$.
Usually, it is assumed that the real currents vanish at $\tau<\tau_0$.  
Substituting this source into the Eqs.~(\ref{eq:E2.37}), (\ref{eq:E2.39}) 
and (\ref{eq:E2.43}), and taking the limit of $\tau_1\rightarrow\tau_0$,
we may verify that the standard prescription for the solution of the
initial data problem,
\begin{eqnarray}
A_{l}(x_1)=\int_{(\tau_2=\tau_0)} d^2{\vec r}_2~d\eta_2~\Delta_{lm}(x_1,x_2) 
{\stackrel{\leftrightarrow}{\partial\over\partial\tau_2}}
A^m(x_2)~~,
\label{eq:E2.48}\end{eqnarray}     
holds with the same propagator $\Delta_{lm}(x_1,x_2) $ that was used to
solve the emission problem. For example, in the limit of
$\tau\rightarrow\tau_0$, the $\eta$-component of the vector 
potential is a sum of three terms, 
\begin{eqnarray} 
A^{(tr)}_{\eta}(\tau_0+0)={i\pi\over 4k_{\bot}^{2}}~\big\{~ 
[R^{(2)}_{1,i\nu}(k_{\bot}\tau_0)H^{(1)}_{i\nu}(k_{\bot}\tau_0)-
R^{(1)}_{1,i\nu}(k_{\bot}\tau_0)H^{(2)}_{i\nu}(k_{\bot}\tau_0)]
[\nu {\bar A}_\phi -k_{\bot}^{2}{\bar A}_\eta] \nonumber \\   
-\tau_0\nu 
[R^{(2)}_{1,i\nu}(k_{\bot}\tau_0) R^{(1)}_{-1,i\nu}(k_{\bot}\tau_0)-
R^{(1)}_{1,i\nu}(k_{\bot}\tau_0) R^{(2)}_{-1,i\nu}(k_{\bot}\tau_0)]
{\bar A'}_\phi ~\big\} ~~,
\label{eq:E2.49}\end{eqnarray}      
\begin{eqnarray} 
A^{(L)}_{\eta}(\tau_0+0)={- \nu\over k_{\bot}^{2}}
\big\{~-s_{1,i\nu}(k_{\bot}\tau_0)
{\bar A}_\phi + \tau_0 Q_{-1,i\nu}(k_{\bot}\tau_0){\bar A'}_\phi   
-{\nu k_{\bot}^{2}\over\tau_0} s_{-1,i\nu}(k_{\bot}\tau_0){\bar A}_\eta
+{\nu \over\tau_0} Q_{1,i\nu}(k_{\bot}\tau_0){\bar A'}_\eta~\big\}~~,
\label{eq:E2.50}\end{eqnarray}              
\begin{eqnarray} 
A^{(inst)}_{\eta}(\tau_0+0)= {\nu\over k_{\bot}^{2}}
Q_{1,i\nu}(k_{\bot}\tau_0) [\tau_0{\bar A'}_\phi 
+{\nu \over\tau_0}{\bar A'}_\eta ]~~,
\label{eq:E2.51}\end{eqnarray}
where we have denoted: ${\bar A}_\phi = k_x{\bar A}_x +k_y{\bar A}_y $.    
Here, Eq.~(\ref{eq:E2.51}) follows from Eq.~(\ref{eq:E2.43}) and takes care 
of the consistency between the charge density at the moment $\tau_0$ and the
initial data for the gauge field. Using relations (\ref{eq:E2.30}) and
(\ref{eq:E2.42}) and adding up Eqs.~(\ref{eq:E2.49})--(\ref{eq:E2.51})
we come to a desired identity, $A_{\eta}(\tau_0+0)= {\bar A}_\eta $.

When the initial data ${\bar A}_m({\vec r},\eta)$   and  ${\bar A'}_m
({\vec r},\eta)$ correspond to the free radiation field, then only the part
$\Delta^{(tr)}_{lm}(x_1,x_2)$ of the full propagator ``works'' here, and
only Eq.~(\ref{eq:E2.49}) may be retained. The other two equations acquire
status of the constraints  imposed on the initial data. Since
the current is absent, we have $A^{(L)} =0$ on the left hand side of the
Eqs.~(\ref{eq:E2.50}). Then the right hand side confirms that the
kernel  ${\cal K}$ is orthogonal to the free radiation field modes.
Since the charge density $\rho$ vanishes, we have 
$A^{(inst)} =0$, which is equivalent to the Gauss law for the free gauge
field. The two transverse modes already obey these constraints.
This fact provides a reliable footing for canonical
quantization of the free field in the gauge $A^\tau=0$. Indeed, the Riemann
function induces commutation relations for the gauge field. It can be
found via its bilinear decomposition over the physical modes. Thus one
can avoid technical problems of inverting the constraint equations. (See
Sec.~\ref{sec:SN3}.) The longitudinal part of the propagator will be
studied in details in Sec.~\ref{sec:SN5}. 

\subsection{Gluon vertices in the gauge $A^{\tau}=0$} 
\label{subsec:SB24} 
 
The terms proportional to the first and the second powers of the coupling
constant in the classical wave equations may be viewed as the external
current and allow one to define the explicit form of the 3- and 4-gluon
vertices. One should start from the solution of the Maxwell equations,
\begin{eqnarray}
A^{a'}_{k'}(z_1)=\int d^4 x
\Delta^{a'a}_{k'k}(z_1,x) \sqrt{-{\rm g}(x)} {\cal J}^{k}_{a}(x)~~.     
\label{eq:E2.52}\end{eqnarray}      
with the color current of the form
\begin{eqnarray}
\sqrt{-{\rm g}(x)} {\cal J}^{k}_{a}(x)=-gf_{abc}\sqrt{-{\rm g}(x)}
{\rm g}^{kn}(x){\rm g}^{ml}(x) [\partial_m(A^{b}_{l}(x)A^{c}_{n}(x))+
A^{b}_{l}(x) \partial_m A^{c}_{n}(x) + 
A^{b}_{m}(x)\partial_n A^{c}_{l}(x) ] \nonumber \\
-g^2 \sqrt{-{\rm g}(x)} f_{abc}f_{cdh} {\rm g}^{kn}(x){\rm g}^{ml}(x)
A^{b}_{l}(x)A^{d}_{m}(x)A^{h}_{n}(x)~~.     
\label{eq:E2.53}\end{eqnarray} 
In perturbation calculations, every field $A(x)$ in the RHS of this
expression is a part of some correlator $\Delta(x,z_N)$. The components of
the metric depend only on the time $\tau$ while the derivatives affect only
the spacial directions $u^n=({\vec r},\eta)$. Moreover, in these
directions, all the gluon correlators depend only on the differences
of the coordinates and can be rewritten in terms of their spacial Fourier
components. After symmetrization over the outer arguments $z_N$, one
immediately obtains,
\begin{eqnarray}
V^{kln}_{abc}(p_1,p_2,p_3;\tau)=-i\tau f_{abc}~\delta(p_1+p_2+p_3)~ 
[{\rm g}^{ln}(p_2-p_3)^k+
{\rm g}^{nk}(p_3-p_1)^l+{\rm g}^{kl}(p_1-p_2)^n]~~,     
\label{eq:E2.54}\end{eqnarray}        
where $p^n={\rm g}^{nk}p_k$, and the components of the momentum in the 
curvilinear coordinates are equal to $p_k=(p_x,p_y,\nu)$.
The four-gluon vertex has no derivatives and is the same as usually.

\section{Quantization}   
\label{sec:SN3} 

The second quantization of the field has several practical goals. 
We would like to have an expansion of the  operator of the free
gluon field like
\begin{equation} 
A_i (x)=\sum_{\lambda=1,2} \int d^2 {\vec k} d\nu [ c_{\lambda}(\nu,{\vec k})
V^{(\lambda)}_{\nu {\vec k};i}(x) + c^{\dag}_{\lambda}(\nu,{\vec k})  
V^{(\lambda)\ast }_{\nu {\vec k};i}(x)]~~,              
\label{eq:E3.1}\end{equation}        
with the creation and annihilation operators which obey the commutation
relations
\begin{eqnarray}
[c_{\lambda}(\nu,{\vec k}),c^{\dag}_{\lambda'}(\nu',{\vec k}')] 
=\delta_{\lambda\lambda'}\delta(\nu-\nu')\delta({\vec k}-{\vec k}'),\;\;\;
[c_{\lambda}(\nu,{\vec k}),c_{\lambda'}(\nu',{\vec k}')]= 
[c^{\dag}_{\lambda}(\nu,{\vec k}),
c^{\dag}_{\lambda'}(\nu',{\vec k}')]=0~~. 
\label{eq:E3.2}\end{eqnarray} 
Once obtained, commutation relations (\ref{eq:E3.2}) allow one to find
various correlators of the free gluon field as the averages of the binary
operator products over the state of the perturbative vacuum and express
them  via the solutions $V^{(1)}_{\nu {\vec k};i}(x)$ and  
$V^{(2)}_{\nu{\vec k};i}(x)$. For example, the  Wightman functions,
\begin{eqnarray}
i\Delta_{10,ij}(x,y)= \langle 0|  A_i(x) A_j(y) | 0 \rangle =
\sum_{\lambda=1,2}\int d\nu d^2{\vec k}
V^{(\lambda)}_{\nu {\vec k};i}(x) V^{(\lambda)\ast }_{\nu {\vec k};i}(y)
=i\Delta_{01,ji}(y,x)~~,
\label{eq:E3.3}\end{eqnarray} 
serve as the projectors onto the space of the on-mass-shell  gluons and
should be known explicitly in order to have a good  definition for the
production rate of the  gluons in the final states.
With these two Wightman functions at hand, one immediately obtains the
expression for the commutator of the free field operators,
\begin{eqnarray}
\Delta_{0,ij}(x,y)= -i \langle 0|[  A_i(x), A_j(y)] | 0 \rangle =
  \Delta_{10,ij}(x,y)-\Delta_{01,ij}(x,y)~,
\label{eq:E3.3a}\end{eqnarray}    
which should coincide with the Riemann function of the homogeneous field
equations.  The program of secondary quantization do not reveal any
technical problems if we give preference to the holomorphic quantization
which starts commutation relations  (\ref{eq:E3.2}) for the Fock
operators. However, if we prefer to start with the canonical commutation
relations for the field coordinates and momenta, then one should {\em
postulate} them and derive (\ref{eq:E3.2}) as the consequence.
 
The way to obtain  the canonical commutation relations in cases of the
scalar  and the spinor fields  is quite straightforward. For the vector
gauge  field we meet a well known problem, an excess of the number of the
components of the vector field  over the number of the physical degrees of
freedom. For example, in the  so-called radiation gauge, $A^0=0$ and
${\rm div} {\bf A}=0 $, we write the canonical commutation relations 
in the following form \cite{Bjorken},
\begin{eqnarray}       
\big[ A_{i} ({\bf x},t), E_{j} ({\bf y},t) \big]=
\delta_{ij}^{tr} ({\bf x}-{\bf y})
=\int {d^3 {\bf k}\over (2\pi )^3 } \bigg( \delta_{ij}-
{ k_i k_j \over {\bf k}^2 } \bigg) e^{-ik(x-y)} ,  \nonumber    \\
\big[ A_{i} ({\bf x},t), A_{j} ({\bf y},t) \big]  = 
\big[ E_{i} ({\bf x},t), E_{j} ({\bf y},t) \big]=0~, 
\label{eq:E3.4}\end{eqnarray} 
thus eliminating  the longitudinally polarized photons from the dynamical
degrees of freedom. The function $\delta_{ij}^{tr}$  plays a role as the
unit operator in the space of the physical states. Here, $i,j=1,2,3$ and 
the number of relations postulated by equations (\ref{eq:E3.4}) apparently
exceeds the actual number required by the count of the independent degrees
of freedom, $\lambda=1,2$, of the free gauge field. The Fourier transform
of the function $\delta_{ij}^{tr}$ is easily guessed because the basis  
of the plane-wave  solutions is very simple \cite{Bjorken}, and it can be
obtained rigorously by solving the system of constraint equations
\cite{Weinberg,Tyutin}. A similar guess or procedure in our  case is not
so obvious. We have the  gauge condition $A^\tau=0$ as the primary
constraint and the Gauss law as the secondary one. The latter can be
resolved in a way which allows one to exclude the $\eta$-components of the
potential and the electric field  from the set of independent canonical
variables. Thus, only $x$- and $y$-components are subject for the canonical
commutation relations. To resolve the constraints, one anyway needs the
integral operators with the kernels built from the solutions of the
Maxwell equations in the gauge $A^\tau =0$. Therefore we shall proceed in
two steps. In section ~\ref{subsec:SB31} we shall sketch the results for
the Wightman functions  (\ref{eq:E3.3}). These, will be used for the
explicit calculation of the free field commutator (\ref{eq:E3.3a}) in
section ~\ref{subsec:SB32} and for the study of its causal behavior.

\subsection{Gluon correlators in the gauge $A^{\tau}=0$} 
\label{subsec:SB31}

Here, we shall write down components of the field correlator 
$\Delta_{10,ij}(x,y)$ in the curvilinear coordinates 
$u=(\tau, \eta,{\vec r})$. We shall denote their covariant components as
$\Delta_{10,ik}(u_1,u_2)$. Later we shall transform them to the standard
Minkowsky coordinates and find the correlators of the temporal axial and
the null-plane gauges as their limits in the central rapidity region and
in the vicinity of  the null-planes, respectively. The  most convenient 
for this purpose basis consists of the transverse modes  $v^{(\lambda)}$. 
The mode $v^{(1)}$ gives the following contribution to the
correlator $\Delta_{10,ik}$:
\begin{eqnarray}
i \Delta^{(1)}_{10,rs}(1,2)= \int_{-\infty}^{\infty} {d\theta\over 2}
\int {d^2{\vec k}\over (2\pi)^3} 
{\epsilon_{ru}\epsilon_{sv} k_u k_v \over k_{\bot}^{2}}   
e^{i{\vec k}({\vec r}_1-{\vec r}_2)}
e^{-ik_{\bot}\tau_1 \cosh (\theta-\eta_1)+ik_{\bot}\tau_2 
\cosh (\theta-\eta_2)}~. 
\label{eq:E3.5} 
\end{eqnarray}           
Realising that $d\theta/2=dk^3/2k^0$, we recognize a standard representation
of this part of the correlator in terms of the on-mass-shell
plane waves decomposition.

The second part of the correlator is determined by the mode $v^{(2)}$
and has the following components:                                            
\begin{eqnarray}
\Delta^{(2)}_{10,rs}(1,2)= -i\int_{-\infty}^{\infty} {d\theta\over 2}
\int {d^2{\vec k}\over (2\pi)^3} 
{k_r k_s \over k_{\bot}^{2}}   
e^{i{\vec k}({\vec r}_1-{\vec r}_2)} 
f_1(\theta,\tau_1 ,\eta_1)f_{1}^{*}(\theta,\tau_2 ,\eta_2)~,   
\label{eq:E3.6} 
\end{eqnarray}         
\begin{eqnarray}
\Delta^{(2)}_{10,r\eta}(1,2)= -i \int_{-\infty}^{\infty} {d\theta\over 2}
\int {d^2{\vec k}\over (2\pi)^3}   
e^{i{\vec k}({\vec r}_1-{\vec r}_2)}{k_r\over k_{\bot}^2}
 f_1(\theta,\tau_1 ,\eta_1)f_{2}^{*}(\theta,\tau_2 ,\eta_2)
=i\Delta^{(2)}_{10,\eta r}(2,1)~, 
\label{eq:E3.7} 
\end{eqnarray}    
\begin{eqnarray}
\Delta^{(2)}_{10,\eta\eta}(1,2)= -i\int_{-\infty}^{\infty}{d\theta\over 2}
\int {d^2{\vec k}\over (2\pi)^3}     
{e^{i{\vec k}({\vec r}_1-{\vec r}_2)}\over k_{\bot}^{2}} 
f_2(\theta,\tau_1 ,\eta_1)f_{2}^{*}(\theta,\tau_2 ,\eta_2)~.   
\label{eq:E3.8} 
\end{eqnarray} 
One may easily see that all components of $\Delta_{10}(1,2)$
vanish when either $\tau_1$ or $\tau_2$ go to zero.         

\subsection{Causal properties of the field commutators 
            in the gauge $A^{\tau}=0$} 
\label{subsec:SB32}

 Causal properties of the radiation field commutator may be studied 
starting from the representation (\ref{eq:E3.3a}). Using Eqs. (\ref{eq:E3.5})
and (\ref{eq:E3.6}) we may conveniently wright contribution of two
transverse modes in the following form:
\begin{eqnarray}
i \Delta^{(1)}_{0,rs}(1,2)= -i \int {d^2{\vec k}\over (2\pi)^3} 
{\epsilon_{ru}\epsilon_{sv} k_u k_v \over k_{\bot}^{2}}  
e^{i{\vec k}{\vec r}}  \int_{-\infty}^{\infty} d\theta\sin k_\bot \Phi~,
\label{eq:E3.9} 
\end{eqnarray}     
\begin{eqnarray} 
i\Delta^{(2)}_{0,rs}(1,2)= -i \int {d^2{\vec k}\over (2\pi)^3} 
{k_r k_s \over k_{\bot}^{2}} e^{i{\vec k}{\vec r}} 
\int_{-\infty}^{\infty} d\theta \bigg[ 1-
{\cosh 2\eta \over \sinh^2 \theta -\cosh^2 \eta} \bigg]
 (\sin k_\bot \Phi 
-\sin k_\bot \Phi_1 +\sin k_\bot \Phi_2 )~,  
\label{eq:E3.10} 
\end{eqnarray}        
where we have introduced the following  notation:~
$2\eta=\eta_1-\eta_2~,~~{\vec r}={\vec r}_1 - {\vec r}_2~,~~ 
\Phi_i =\tau_1\cosh (\theta -\eta_i)~,~~ \Phi=\Phi_1-\Phi_2 ~$.
The sum of (\ref{eq:E3.9}) and (\ref{eq:E3.10}) can be rearranged to the
following form,
\begin{eqnarray} 
i\Delta_{0,rs}(1,2)= i\int {d^2{\vec k}~d\theta \over (2\pi)^3} 
e^{i{\vec k}{\vec r}} \bigg[ - \delta_{rs} \sin k_\bot \Phi +
{k_r k_s \over k_{\bot}^{2}}[ \sin k_\bot \Phi_1- 
\sin k_\bot \Phi_2]\nonumber \\ 
+ k_r k_s \cosh(\eta_1-\eta_2)\int_{0}^{\tau_1}d\tau' 
\int_{0}^{\tau_2}d\tau''~\sin [ k_\bot\tau'\cosh(\theta -\eta)-
k_\bot\tau'\cosh(\theta + \eta)] \bigg]~~. 
\label{eq:E3.10a} 
\end{eqnarray}                                         
Joining the integration $d^2{\vec k}~d\theta$ into the three dimensional
integration ~$d^3 {\bf k} / | {\bf k}|$~  in the Cartesian coordinates,
the first integral in (\ref{eq:E3.10a}), 
\begin{eqnarray} 
D_{0}(1,2)= \int {d^2{\vec k}~d\theta \over (2\pi)^3} 
e^{i{\vec k}{\vec r}} \sin k_\bot \Phi ={{\rm sign}(t_1-t_2) \over 2\pi}
\delta [(t_1-t_2)^2-({\bf r}_1-{\bf r}_2)^2]~, 
\label{eq:E3.11} 
\end{eqnarray}    
is easy to calculate, and to recognize it as the commutator of the massless
scalar field. It differs from zero only if the line between the
points $x_1$ and $x_2$ has the light-like direction.
In this way we integrate the first and the third terms in the integrand of
the Eq.~(\ref{eq:E3.10a}). To reduce two integrals in the second term 
to the same type, we must exclude the factor $1/k_{\bot}^{2}$ using
the fundamental solution of the two-dimensional Laplace operator,
\begin{eqnarray} 
{k_r k_s \over k_{\bot}^{2}} e^{i{\vec k}{\vec r}}=
\partial_r\partial_s \int {d^2{\vec \xi} \over 2\pi} 
\ln |{\vec \xi}-{\vec r}|e^{i{\vec k}{\vec \xi}}~~.
\label{eq:E3.12} 
\end{eqnarray}    
After that we arrive at the final result,
\begin{eqnarray} 
\Delta_{0,rs}(1,2)= - \delta_{rs} D_{0}(1,2) -
\cosh(\eta_1-\eta_2)\partial_r\partial_s   \int_{0}^{\tau_1}d\tau_1 
\int_{0}^{\tau_2}d\tau_2  D_{0}(1,2) \nonumber \\
+ \partial_r\partial_s \int {d^2{\vec \xi} \over (2\pi)^2}
\ln |{\vec \xi}-{\vec r}| [\delta (\tau_{1}^{2}-{\vec \xi}^{2})-
\delta (\tau_{2}^{2}-{\vec\xi}^{2})]~~. 
\label{eq:E3.13} 
\end{eqnarray}                                         
From this form it immediately follows that the commutator of the 
potentials vanishes at $\tau_1=\tau_2$.  Even more strong result takes
place for the commutator of the two electric fields,
\begin{eqnarray}
 [ E_r(1), E_s(2)] = 
{\partial^2 \over \partial \tau_1 \partial \tau_2 }
i\Delta_{0,rs}(1,2)= 
\bigg[ - \delta_{rs} {\partial^2 \over \partial \tau_1 \partial \tau_2 }
- \cosh(\eta_1-\eta_2) 
{\partial^2 \over \partial x^r \partial x^s } \bigg]~iD_{0}(1,2)~~.
\label{eq:E3.14}\end{eqnarray}                                
This commutator vanishes everywhere except for the light cone, in full
compliance with the microcausality principle for the electric field
which is an observable. However, this does not happens for the commutator
of the potentials since they are defined nonlocally.  It does not
vanish neither at space-like  nor at space-like separation,
because the line of integration which recovers the potential at the point
$x_2$, in general, intersects ({\em e.g.} at some point $x_3$) with the light 
cone which has its vertex at the point $x_1$, and the commutator of the
electric fields at the points $x_1$ and $x_3$ is not zero.

Similar results take place for the commutator of the $\eta$-components of
the potential and the electric field. The field commutator,
\begin{eqnarray}
 [ E_\eta (1), E_\eta(2)] = 
{\partial^2 \over \partial \tau_1 \partial \tau_2 }
i\Delta_{0,\eta\eta}(1,2)= -i \nabla_{\bot}^{2} D_{0}(1,2)~~,
\label{eq:E3.15}\end{eqnarray}                                
is entirely causal, while the commutator of the potentials,
\begin{eqnarray}
 [ A_\eta (1), A_\eta (2)]=i\Delta_{0,\eta\eta}(1,2)=-i \nabla_{\bot}^{2} 
\int_{0}^{\tau_1}\tau_1d\tau_1\int_{0}^{\tau_2}\tau_2d\tau_2D_{0}(1,2)~~,
\label{eq:E3.16}\end{eqnarray}    
does not vanishes at space-like distances, except for $\tau_1=\tau_2$.
Finally, the formally designed commutator between the $r$-- and 
$\eta$--components of the electric field, the two observables,
\begin{eqnarray}
 [ E_r (1), E_\eta(2)] = 
{\partial^2 \over \partial \tau_1 \partial \tau_2 } 
i\Delta_{0,r\eta}(1,2)= - {\partial^2 \over \partial x^r \partial \eta } 
\bigg( {\tau_2\over\tau_1}D_{0}(1,2)\bigg)~~,
\label{eq:E3.17}\end{eqnarray}      
is entirely confined to the light cone, while the commutator of the
potentials, which are not the observables,
\begin{eqnarray}
 [ A_r (1), A_\eta(2)] = i\Delta_{0,r\eta}(1,2)=
-\int_{0}^{\tau_1} {d \tau_1\over\tau_1} \int_{0}^{\tau_2}\tau_2 d \tau_2  
 {\partial^2 \over \partial x^r \partial \eta } D_{0}(1,2)~~,
\label{eq:E3.18}\end{eqnarray}      
does not vanish at the space-like distance, even at $\tau_1=\tau_2$.
This result, however, is not a subject for any concern since the potentials
are defined nonlocally and commutation relations for electric and magnetic
[cf. (\ref{eq:E2.15A})] fields are reproduced correctly. Moreover, we have
argued above that the $\eta$-components of $A$ and $E$ are not the
canonical variables since the constraints express them via $x$- and
$y$-components.

The ``acausal'' behavior of the Riemann function ~$\Delta_{0}^{\mu\nu}(1,2)$  
may cause doubts if the gauge $A^\tau =0$ allows for meaningful retarded
and advanced Green functions which, by causality, should vanish at
space-like distances. Fortunately, this anomalous behavior appears only
for the gauge--variant potential; the response functions for observable
electric and magnetic fields are causal. This can be easily seen,  {\em
e.g.}, from Eqs. (\ref{eq:E2.19}), (\ref{eq:E2.23})  and (\ref{eq:E2.24}),
the usual inhomogeneous relativistic wave equations for various physical
components of the field strengths ${\cal E}$ and  ${\cal B}$.

\subsection{Canonical commutation relations in the gauge $A^{\tau}=0$} 
  \label{subsec:SB33}

A proof of the commutation relations (\ref{eq:E3.2}) for the Fock 
operators follows the
standard guidelines \cite{Bjorken}. The creation and annihilation
operators are {\em defined} via relations,
\begin{eqnarray}
c_{\lambda}(\nu,{\vec k})= (V^{(\lambda)}_{\nu {\vec k}},A)=
i{\bf g}^{ij}\int d^3{\bf x}
[ V^{(\lambda)\ast }_{\nu {\vec k};j}(x){\stackrel{\bullet} A}_i({\bf x},\tau)
-{\stackrel{\bullet} V}^{(\lambda)\ast }_{\nu {\vec k};j}(x)
                            A_i({\bf x},\tau)]~, \nonumber \\
c^{\dag}_{\lambda}(\nu,{\vec k}) =(A,V^{(\lambda)}_{\nu {\vec k}})=
i{\bf g}^{ij}\int d^3{\bf x}
[A_i({\bf x},\tau){\stackrel{\bullet} V}^{(\lambda) }_{\nu {\vec k};j}(x)
-{\stackrel{\bullet} A}_i({\bf x},\tau)V^{(\lambda) }_{\nu {\vec k};j}(x)]~~. 
\label{eq:E3.19}\end{eqnarray}  
The latter result in the following expression for the commutator,
\begin{eqnarray}
[c_{\lambda}(\nu,{\vec k}),c^{\dag}_{\lambda'}(\nu',{\vec k'})] =
\int d^3{\bf x}d^3{\bf y} {\rm g}^{ij}(x){\rm g}^{lm}(y)~
\big\{ [A_i({\bf x},\tau),{\stackrel{\bullet} A}_l({\bf y},\tau)]~
\big({\stackrel{\bullet} V}^{(\lambda)\ast }_{\nu {\vec k};j}(x)
V^{(\lambda')}_{\nu'{\vec k}';n}(y)
-{\stackrel{\bullet} V}^{(\lambda')\ast}_{\nu' {\vec k}';j}(x)  
V^{(\lambda)}_{\nu{\vec k};n}(y)\big) \nonumber \\ 
+[A_i({\bf x},\tau),A_l({\bf y},\tau)]~
{\stackrel{\bullet} V}^{(\lambda)\ast }_{\nu {\vec k};j}(x)
{\stackrel{\bullet} V}^{(\lambda')}_{\nu'{\vec k}';n}(y) +
[{\stackrel{\bullet} A}_i({\bf x},\tau),{\stackrel{\bullet} A}_l({\bf y},\tau)]~
 V^{(\lambda')\ast}_{\nu' {\vec k}';j}(x)
V^{(\lambda)}_{\nu{\vec k};n}(y) \big\} ~~. 
\label{eq:E3.20}\end{eqnarray} 
Most of the terms in the second line vanish due to the commutation
relations. Next, we rely on the following guess about the form of the 
commutator,    
\begin{eqnarray}
[A_i(x), A_j(y)]=\sum_{\lambda=1,2}\int d\nu ~d^2{\vec k}~
\big( V^{(\lambda)}_{\nu{\vec k};i}(x)V^{(\lambda)\ast }_{\nu{\vec k};j}(y)-
V^{(\lambda)\ast}_{\nu{\vec k};i}(x)
V^{(\lambda)}_{\nu {\vec k};i}(y)\big) ~~,
\label{eq:E3.22}\end{eqnarray} 
which leads to the proper equal-time commutation relations for the
independent canonical variables.
Finally, explicitly using the orthogonality relations for the eigenmodes
$V^{(\lambda)}$, we immediately obtain the commutation relations 
(\ref{eq:E3.2}).  

\section{Longitudinal propagator and static fields} 
\label{sec:SN5}
         
In this section we shall find the explicit expressions for the kernels
(\ref{eq:E2.40}) and (\ref{eq:E2.44}) which represent the longitudinal and 
instantaneous components of the gauge field produced by the ``external''
current $j^\mu$. The calculations are lengthy and their details are
adduced in Appendix~3. Here, we present only the final answers. 

The components of the longitudinal propagator
$\Delta^{(L)}_{lm}(\tau_2,{\vec r},\eta)$ are already obtained in the form 
of the three-dimensional integrals (\ref{eq:E2.40}). $\Delta^{(L)}$
depends on the differences of the curvilinear spatial coordinates,
${\vec r}={\vec r}_1-{\vec r}_2$ and $\eta=\eta_1-\eta_2$, but {\em not}
on the difference of the temporal arguments $\tau_1$ of the field and
$\tau_2$ of the source.  Introducing the shorthand notation for the 
distance in the $(xy)$--plane, $r_{\bot}=|{\vec r}|$, and for the full
distance $R_2=R(\tau_2)=[({\vec r}_1-{\vec r}_2)^2+
\tau_{2}^{2}\sinh^2(\eta_1-\eta_2)]^{1/2}$ 
between the two points of the surface $\tau_2=const$, we obtain:
\begin{eqnarray} 
\Delta^{(L)}_{rs}= -{\theta(\tau_2-r_{\bot}) \over 4\pi}  
{\partial^2\over\partial x^r\partial x^s}
\bigg[L_2~\coth |\eta| -\ln{\tau_2\over r_{\bot}}\bigg]~,  \nonumber\\
\Delta^{(L)}_{\eta s}= -{\theta(\tau_2-r_{\bot}) \over 4\pi}
{\partial^2\over\partial\eta\partial x^s} (L_2~\coth |\eta|)~,~~~ 
\Delta^{(L)}_{r \eta}= -{\theta(\tau_2-r_{\bot}) \over 4\pi}
{1\over \cosh\eta} {\partial^2\over\partial x^r\partial\eta}
\bigg({L_2 \over\sinh |\eta|}\bigg)~,  \nonumber\\
\Delta^{(L)}_{\eta\eta}= {\tau_{2}^{2}\over 2}\delta({\vec r})\delta(\eta)
+{\theta(\tau_2-r_{\bot}) \over 4\pi}
\bigg[ 2{\eta\coth\eta-1 \over\sinh^2\eta}  +{\nabla_{\bot}^{2}\over 2}
\bigg( -{r_{\bot}\cosh\eta \over\sinh^3 |\eta|}~L_2  +
{\tau_2 R_2 \cosh\eta \over \sinh^2\eta}\bigg)\bigg]~~.
\label{eq:E5.1}\end{eqnarray}                  
where $~L_2=L(\tau_2)=\ln [(\tau_2\sinh |\eta|+R_2)/r_{\bot}]$.  After the 
derivatives are evaluated, most of the logarithms here vanish: 
\begin{eqnarray} 
\Delta^{(L)}_{rs}= -{\theta(\tau_2-r_{\bot}) \over 4\pi}
\bigg[{1\over r_{\bot}^{2}} \bigg(1-{\tau_2\cosh\eta\over R_2}\bigg)
\bigg(\delta_{rs}- {2 x^r x^s\over r_{\bot}^{2}}\bigg)  
-{2 x^r x^s\over r_{\bot}^{2}} 
{\tau_2\cosh\eta\over R_{2}^{3}}\bigg]~,\nonumber\\
\Delta^{(L)}_{\eta s}= -{\theta(\tau_2-r_{\bot}) \over 4\pi}
{x^s\over r_{\bot}^{2}}{\tau_2\sinh\eta\over R_2}
{\tau_{2}^{2}-r_{\bot}^{2}\over R_{2}^{2}}~,~~~ 
\Delta^{(L)}_{r \eta}= -{\theta(\tau_2-r_{\bot}) \over 4\pi}
{x^r \over r_{\bot}^{2}}{\tau_{2}^{3}\sinh\eta\over R_{2}^{3}}~,  \nonumber\\
\Delta^{(L)}_{\eta\eta}= {\tau_{2}^{2}\over 2}\delta({\vec r})\delta(\eta)
+{\theta(\tau_2-r_{\bot}) \over 4\pi}
\bigg[ 2{\eta\coth\eta-1 \over\sinh^2\eta}  +
{\tau_2  \cosh\eta \over R_2 \sinh^2\eta}
\bigg( 3 -{r_{\bot}^{2}\over R_{2}^{2}}\bigg)  -
{2 \cosh\eta \over \sinh^3|\eta|}~L_2 \bigg]~~.
\label{eq:E5.1a}\end{eqnarray}                  
By  examination of Eq.~(\ref{eq:E2.39}), one may see that after replacement
of $\tau_2$ by $\tau_1$ the same kernel, 
$\Delta^{(L)}_{lm}(\tau_1,{\vec r}, \eta)$, determines the components
$ E_{m}^{(L)}(\tau_1)$ of the longitudinal part of the electric field via 
the components $j^m(\tau_1)$ of the current at the same time.

Similar calculations have led to the expression (\ref{eq:E2.18}) for the
electric field of the static Coulomb source. The  kernel ${\cal K}_m$  of
the instantaneous potential  has the following components,
\begin{eqnarray} 
{\cal K}_r = -{\theta(\tau_1-r_{\bot}) \over 4\pi} \cosh\eta
{\partial L_1\over\partial x^r} = {\theta(\tau_1-r_{\bot}) \over 4\pi} 
{x^r \over r_{\bot}^{2}}
{\tau_2~\cosh\eta~\sinh |\eta|\over R_1}~,\nonumber \\
{\cal K}_\eta = -{\theta(\tau_1-r_{\bot}) \over 4\pi}
\bigg( {\tau_1\over R_1 \sinh |\eta|} + 
{|\eta|-\tanh|\eta| \over \sinh^2\eta}-L_1 \bigg)~~,
\label{eq:E5.2}\end{eqnarray} 
where $R_1=R(\tau_1)$ and $L_1=L(\tau_1)$. 
These propagators do not respect the light cone, but have a remarkable 
property that the longitudinal fields at the surface of the constant
proper time $\tau$ do not exist at the distances $r_\bot$ from their sources  
that exceed $\tau$. This establish the upper limit for the possible
dynamical correlations between the longitudinal fields in the $(xy)$-plane.     
\bigskip

\section{Gluon correlators in the central rapidity region and 
         near the light wedge} 
\label{sec:SN4}         
                         
Our next step is to compare the correlators  of the gauge $A^\tau=0$ with
the similar correlators in  three other gauges, $A^0=0$, $A^+=0$ and
$A^-=0$.  We shall start with the simplest on-mass-shell Wightman 
function $\Delta_{10}^{\mu\nu}$. This type correlators, 
$\Delta_{01}^{\mu\nu}$, $\Delta_{0}^{\mu\nu}$ and $\Delta_{1}^{\mu\nu}$
share the same polarization sum of the free gauge field. They correspond
to the densities of the final states of the  radiation field and are
important for various calculations.  The same polarization  sum appears in
expressions for the transverse part of the propagators,
$\Delta_{ret}^{\mu\nu}$, $\Delta_{adv}^{\mu\nu}$, $\Delta_{00}^{\mu\nu}$ 
and $\Delta_{11}^{\mu\nu}$. For our immediate purpose we shall include the
projector $d^{\mu\nu}$ of the gauge $A^\tau=0$ into the formal Fourier
representation,

\begin{eqnarray}
i D^{\mu\nu}_{10}(x_1,x_2)= 
\int {d^3 k\over (2\pi)^3 2k^0}  d^{\mu\nu}(k; x_1,x_2) 
e^{-ik(x_1-x_2)} 
\label{eq:E4.1} 
\end{eqnarray}        
with the ``extraneous'' dependence of the Fourier transform on the
time and spacial coordinates.  This dependence disappear 
in some important limits.  Therefore, we discover the domains
where the wedge dynamic simplifies and describe the
processes which are approximately homogeneous in space and time.
These domains are: (i) the central rapidity region, $\eta_{1,2}\ll 1$ (or
$x^{3}_{1,2}\sim 0$), where the projector in the integrand of
Eq.~(\ref{eq:E4.1}) is
\begin{eqnarray}
 d^{\mu\nu}(k,u)= -{\rm g}^{\mu\nu} +
{k^\mu u^{\nu}+u^\mu k^{\nu} \over  k u } - {k^\mu k^{\nu} \over  (k u)^2 } 
\label{eq:E4.2} 
\end{eqnarray}       
with the gauge-fixing vector, $u^\mu =(1,0,0,0)$, which approximately 
coincide with the local normal to the hypersurface $\tau =const$;
and (ii), the vicinities of two null-planes, $\eta\rightarrow\pm\infty$
(or $x^\mp\rightarrow 0$) where 
\begin{eqnarray}
 d^{\mu\nu}(n_{\pm},k)= -{\rm g}^{\mu\nu} +
{k^\mu n^{\nu}_{\pm}+k^\nu n^{\mu}_{\pm} \over  (k n_{\pm}) }
\label{eq:E4.3} 
\end{eqnarray}       
with the null-plane vectors ~$n^{\mu}_{\pm}=(1,0,0,\mp 1)~$. 

Eqs.~(\ref{eq:E3.5})-(\ref{eq:E3.8}) almost fit our needs.
In all three cases, $x^3\rightarrow 0$, $k^0x^0\gg 1$,  as well as 
$x^-\rightarrow 0$,  $k^-x^+ \gg 1$  and 
$x^+\rightarrow 0$,  $k^+x^- \gg 1$,  the functions $f_1$ and $f_2$
can be approximated by the following expressions
\begin{eqnarray}
f_{1} \approx 
 i~ \tanh (\theta-\eta) e^{-ik_{\bot}\tau \cosh (\theta-\eta)
+i{\vec k}{\vec r} } =
{k^0x^3-k^3x^0 \over k^0x^0-k^3x^3}e^{-ikx}=
{k^+x^- -k^-x^+ \over k^+ x^- +k^- x^+}e^{-ikx}~~. 
\label{eq:E4.4}
\end{eqnarray}         
\begin{eqnarray}
f_{2} \approx 
 ik_{\bot}\tau~ {e^{-ik_{\bot}\tau \cosh (\theta-\eta)
+i{\vec k}{\vec r} } \over \cosh (\theta-\eta)}=
ik_{\bot}^{2}\tau^2 {e^{-ikx}\over k^0x^0-k^3x^3}=
2ik_{\bot}^{2}\tau^2 {e^{-ikx} \over k^+ x^- +k^- x^+}~~. 
\label{eq:E4.4a}
\end{eqnarray}          
(We have omitted the time independent terms in $f_1$ and $f_2$ which
set the potentials of the mode $v^{(2)}$ to zero at $\tau=0$. This
kind of terms would correspond to the residual gauge symmetry and
is not kept in the axial and the null-plane gauges as well.)

Transformation of the correlator $\Delta^{lm}(1,2)$ to the Minkowsky
coordinates is carried out according to the formula,
\begin{equation}
D^{\mu\nu}(x_1,x_2)=a^{\mu}_{i}(x_1){\rm g}^{il}(x_1) 
\Delta_{lm}(u_1,u_2){\rm g}^{mk}(x_2) a^{\nu}_{m}(x_2)  
\label{eq:E4.5}\end{equation}  
where the matrix of the transformation is defined in a standard way,
\begin{equation}
a^{\mu}_{i}(x) = {\partial x^\mu \over \partial u^i},~~~
a^{0}_{\eta}(x) = x^3,~~~a^{3}_{\eta}(x) = x^0,
~~~a^{r}_{s}=\delta^{r}_{s}~~.
\label{eq:E4.6}\end{equation} 
These are the only components of the tensor $a^{\mu}_{i}(x)$ which
participate the transformation.  In this way we obtain
\begin{eqnarray}
D^{00}(1,2)=x^{3}_{1}x^{3}_{2}\Delta^{\eta\eta}(1,2);\;\; 
D^{03}(1,2)=x^{3}_{1}x^{0}_{2}\Delta^{\eta\eta}(1,2) \nonumber \\
D^{30}(1,2)=x^{0}_{1}x^{3}_{2}\Delta^{\eta\eta}(1,2);\;\; 
D^{33}(1,2)=x^{0}_{1}x^{0}_{2}\Delta^{\eta\eta}(1,2), \nonumber \\
D^{0r}(1,2)=x^{3}_{1} \Delta^{\eta r}(1,2);\;\; 
D^{r0}(1,2)=x^{3}_{2} \Delta^{r\eta }(1,2);\nonumber \\ 
D^{3r}(1,2)=x^{0}_{1} \Delta^{\eta r}(1,2);\;\; 
D^{r3}(1,2)=x^{0}_{2}\Delta^{r\eta}(1,2),\nonumber\\
D^{rs}(1,2)=\Delta^{rs}(1,2)~~. 
\label{eq:E4.7}\end{eqnarray}                              
Every additional factor ${\rm g}^{\eta\eta}=\tau^{-2}$ finds a counterpart
which prevents a singular behavior at $\tau =0$. In the above
approximation, the expression for the  $\Delta^{\eta\eta}(x_1,x_2)$
component of the correlator has the form:
\begin{eqnarray}
\Delta^{\eta\eta}(x_1,x_2)= 
\int {d^3 k\over (2\pi)^3 2k^0} {k_{\bot}^{2}e^{-ik(x_1-x_2)}
\over (k^0x_{1}^{0}-k^3x_{1}^{3})(k^0x_{2}^{0}-k^3 x_{2}^{3})}  
=\int {d^3 k\over (2\pi)^3 2k^0} {4 k_{\bot}^{2}e^{-ik(x_1-x_2)}
\over (k^+x_{1}^{-}+k^-x_{1}^{+})(k^+x_{2}^{-}+k^- x_{2}^{+})}~~,  
\label{eq:E4.8} 
\end{eqnarray}  
Therefore, in the limit of $x^{3}_{1,2}\rightarrow 0$ we obtain that
$D^{00},D^{0i}\rightarrow 0$, while $d^{33}(k,u)\rightarrow 
k_{\bot}^{2}/k_{0}^{2}$, thus reproducing the corresponding components
of the gauge $A^0=0$. The other components are reproduced one by one as well, 
and one can expect smooth transition between the gauge of the the wedge 
dynamic and the local temporal axial gauge of the  reference frame  co-moving
with the dense quark-gluon matter created in the collision.
 
In the limits of $x^{\mp}_{1,2}\rightarrow 0$ we obtain that
\begin{eqnarray}
D^{00},D^{03},D^{30},D^{33}\rightarrow {k_{\bot}^{2}\over (k^{-})^{2}}
={k^+ \over k^-}~,~~~ {\rm if}~~x^-\rightarrow 0~~,    
\label{eq:E4.9} 
\end{eqnarray}  
and 
\begin{eqnarray}
D^{00},D^{03},D^{30},D^{33}\rightarrow {k_{\bot}^{2}\over (k^{+})^{2}}
={k^- \over k^+}~,~~~ {\rm if}~~x^+\rightarrow 0~~.    
\label{eq:E4.10} 
\end{eqnarray}  
These limits, after they are found for all components, lead to the well
known  expressions of the projectors $d^{\mu\nu}(n,k)$ in the null-plane
gauge: the gauge $A^{+}=0$ in the vicinity of $x^+=0$, and  $A^{-}=0$ in
the vicinity of $x^-=0$. Therefore we obtained an expected result: in the
limit of the light-cone  propagation the gauge $A^{\tau}=0$ recovers the
null-plane gauges $A^{+}=0$ and $A^{-}=0$. 
 
Some remarks are in order. First, the conception of the structure
functions relies heavily  on the null-plane dynamics which essentially
uses these gauges. For two hadrons (or two nuclei) we have two different
null-plane dynamics which do not share the same Hilbert space of states.
Now we have an important opportunity to describe both nuclei and the 
fields produced in their interaction within the same dynamic and the same 
Hilbert space. Second, one may trace back the origin of the poles $(ku)^{-1}$
in the polarization sums of axial gauges  $(uA)=0$ and see
that they appeared in course of approximation of the less-singular
factor  $[k_\bot\cosh(\theta-\eta)]^{-1}$ in various limits of the propagator
of the gauge $A^\tau=0$.

Further, contrary to a naive expectation to obtain the gauge $A^{+}=0$ at
$x^{-}=0$ and the gauge $A^{-}=0$ at $x^{+}=0$, we obtained them in the
opposite correspondence. First of all, let us notice that the result is
mathematically consistent. Indeed, the gauge condition  $A^{\tau}=0$ 
may be rewritten in the form, 
\begin{eqnarray} 
A^{\tau}={1\over 2}(A^+ e^{-\eta} + A^- e^{\eta} )=0~~. 
\label{eq:E4.11}  \end{eqnarray}     
Thus the limit of $\eta\rightarrow\infty$ ($x^{-}\rightarrow 0$) indeed 
leads to $A^{-}=0$ and the limit of  $\eta\rightarrow - \infty$
($x^{+}\rightarrow 0$) leads to $A^{+}=0$ as the limit gauge conditions.
Recalling that
\begin{eqnarray} 
A^{\eta}={1\over 2}(A^+ e^{-\eta} - A^- e^{\eta} )= A^+
e^{-\eta} = -A^- e^{\eta} ~~, 
\label{eq:E4.12}  \end{eqnarray}  
we immediately realize that in the vicinities of both null-planes the 
tangent component $A^{\eta}=0$. This fact has a very simple geometrical
explanation: the normal and tangent vectors of the null plane are
degenerate. Once  $A^{\tau}=0$, we have $A^{\eta}=0$ and consequently,
$A^{+}=0$ and $A^{-}=0$ at  $\eta\rightarrow\pm\infty$.  This result
naturally follows from the geometry of the system of the surfaces where we
define the field states. These are subject for dynamical evolution in the
direction  which is normal to the hypersurface.         

The perturbative  QCD-evolution for the object which moves in the
$x^+$-direction and experience the deep inelastic scattering,  begins only
after it has passed the  collision point $x^+=x^-=0$. The so-called
pre-collision dynamics cannot start earlier: the colliding hadron and
electron are Lorentz contracted to a small longitudinal size. It is 
important that the gauge field correlators of the null-plane gauges were
obtained as the sums over the  states which emerge from the collision
point. At $\tau\rightarrow 0$ the stationary phases of  all in-wedge
plane-wave modes are uniformly spread along both null-planes. If no
interaction happens,  then the decomposition remains virtual. The partial
waves will keep their coherence and assemble into the incoming  hadron and
the electron travelling in the initial directions. It takes finite time to
distort the phase balance  between the partial waves of the initial
states. Before the finite distortion develops, the wave packet of the
hadron  partially goes  through the  assembling towards the state of
normal localized hadron.  Only a fraction of the hadron interacts with the
structureless electron.  Therefore, the {\em perturbative} QCD-evolution
of the hadronic wave function indeed takes place in the vicinity of the
null-plane  $x^+=0$,  where the limit of the gauge $A^{\tau}=0$ is
$A^{+}=0$.  It is exactly the domain of large negative $\eta$ where  the
interaction which excites the state of the most hard  back scattered quark
is localized. This qualitative picture is in compliance with the well
known fact that the QCD evolution equations are adequate only  when $x$ is
not too close to 1. 

It is expedient to emphasize  similarity of our approach to the classical
spectral analysis of the transient processes and to the standard scheme of
the quantum-mechanical measurement. Though we decompose the  wave packet
of the incoming hadron in terms of the states which are immediately before
and after the collision are spread over the entire surface of the light
wedge, the phases and amplitudes of this decomposition are balanced in
such a way that  before the collision the whole packet is strongly
localized in the vicinity of the null-plane  $x^-=0$. One needs a strong
interaction in the vertex in order to break down this tiny balance and
filter out an unusual field configuration.


\vskip 3cm  

\noindent {\bf ACKNOWLEDGEMENTS}

\bigskip

I am grateful to A. Belopolsky, D. Dyakonov, B. Muller, L. McLerran,
E.Shuryak, E.Surdutovich and R.Venugopalan  for many stimulating discussions.
I much
profited  from intensive discussions during the International Workshop on
Quantum Chromodynamics and Ultrarelativistic Heavy Ion Collisions at the
ECT${^*}$ in Trento, Italy.

This work was supported by the U.S. Department of Energy under Contract 
No. DE--FG02--94ER40831.          

\bigskip   
\bigskip   
\appendix
\renewcommand{\theequation}{A1.\arabic{equation}}
\setcounter{equation}{0}  

\centerline {\bf APPENDIX~1. Modes of the free gauge field}    

\bigskip

Here, we shall obtain the complete set of the one-particle solutions  to
the homogeneous system  of the Maxwell equations with the gauge 
$A^{\tau}=0$, that is Eqs.~(\ref{eq:E2.10})-(\ref{eq:E2.11}). This gauge 
condition explicitly depends on the coordinates, thus introducing
effective non-locality in the path integral that represents the action.
Therefore it becomes impossible to invert the differential operators using
the standard symbolic methods.  The knowledge of the one-particle
solutions  becomes necessary in order to find  the Wightman functions of
the free vector field, to establish the the explicit form of the field
commutators, and to separate  the propagators of the transverse and the
longitudinal fields.
It is natural to look for the solution in the form of the Fourier
transform with respect to the spacial coordinates:
\begin{eqnarray}  
 A_i (x) = \int_{-\infty}^{\infty} d \nu\int d^2 {\vec k} ~ e^{i\nu\eta}
e^{i{\vec k}{\vec r}} ~A_{i} ({\vec k},\nu ,\tau)
 \label{eq:A1}\end{eqnarray} 
Then the system of the second order ordinary 
differential  equations for the Fourier transforms becomes as follows:
\begin{eqnarray}
[\tau\partial_{\tau}^{2}+\partial_{\tau}+
{\nu^2\over \tau} +\tau k_{y}^{2}]A_x({\vec k},\nu ,\tau)
-\tau k_{x} k_{y}A_y ({\vec k},\nu ,\tau)- 
{\nu k_x\over \tau}A_\eta ({\vec k},\nu ,\tau) =0~~,  
\label{eq:A2}\end{eqnarray}         
\begin{eqnarray}
-\tau k_{x} k_{y}A_x ({\vec k},\nu ,\tau)+
[\tau\partial_{\tau}^{2}+\partial_{\tau}+
{\nu^2\over \tau} +\tau k_{x}^{2}]A_y({\vec k},\nu ,\tau)-
{\nu k_y\over \tau}A_\eta ({\vec k},\nu ,\tau) =0~~,
\label{eq:A3}\end{eqnarray}               
\begin{eqnarray}
-{\nu k_x\over \tau}A_x ({\vec k},\nu ,\tau)-
{\nu k_y\over \tau}A_y ({\vec k},\nu ,\tau) +
[{1\over \tau}\partial_{\tau}^{2}-{1\over \tau^2}\partial_{\tau}+
{1\over \tau}k_{\bot}^{2}]A_\eta ({\vec k},\nu ,\tau) =0~~,
\label{eq:A4}\end{eqnarray}         
In this form the system is manifestly symmetric and self-adjoint.
An additional equation of the constraint reads as
\begin{eqnarray}
{\cal C}({\vec k},\nu ,\tau)=
 {1\over \tau}\nu\partial_{\tau}A_\eta+
\tau\partial_{\tau} [k_x A_x({\vec k},\nu ,\tau) 
+k_{y} A_y({\vec k},\nu ,\tau)] = 0~~.
\label{eq:A5}\end{eqnarray}          
Let us rewrite the homogeneous system of the Maxwell equations
in terms of the variables 
\begin{eqnarray}   
\Phi= \partial_{x}A_x +\partial_{y} A_y ,\;\;\; 
\Psi= \partial_{y}A_x -\partial_{x} A_y ,\;\;\;  {\rm and} \;\; 
{\sf A}=A_\eta .
\label{eq:A6}\end{eqnarray}    
One immediately sees that the equation for the Fourier component
$\Psi({\vec k},\nu ,\tau)$ of the longitudinal magnetic field
$\Psi(x)$ decouple:
\begin{eqnarray}
[\partial_{\tau}^{2}+{1\over \tau}\partial_{\tau}+
{\nu^2 \over \tau^2} +k_{\bot}^{2}]\Psi({\vec k},\nu ,\tau)=0
\label{eq:A7}\end{eqnarray}       
Then the other two equations of motion  take shape
\begin{eqnarray}  
[\tau^2 \partial_{\tau}^{2}+\tau\partial_{\tau}+
\nu^2 ]\Phi_{{\vec k},\nu}(\tau)-i\nu k_{\bot}^2 
{\sf A}_{{\vec k},\nu}(\tau) =0~~,
\label{eq:A8}\end{eqnarray}        
\begin{eqnarray}  
[\partial_{\tau}^{2}-{1\over\tau}\partial_{\tau}+
k_{\bot}^{2}] {\sf A}_{{\vec k},\nu}(\tau) 
+i\nu \Phi_{{\vec k},\nu}(\tau) =0~~, 
\label{eq:A9}\end{eqnarray}  
The additional constraint equation can be conveniently rewritten as     
\begin{eqnarray}
{\cal C}({\vec k},\nu ,\tau)=
 {i\nu\over \tau} \partial_\tau {\sf A}_{{\vec k},\nu}(\tau) 
+\tau\partial_\tau \Phi_{{\vec k},\nu}(\tau) = 0~~.
\label{eq:A10}\end{eqnarray} 
This is an independent equation. However, the conservation of the
constraint along the Hamiltonian time $\tau$ is a consequence of
the equations of motion, and it {\it can} be employed  to obtain
the independent equations for the components of the vector field.        
This is easily done in terms of the auxiliary functions,
\begin{eqnarray}
\varphi_{{\vec k},\nu}(\tau)=\tau{\stackrel{\bullet} \Phi}_{{\vec k},\nu}(\tau)\;\;\; 
{\rm and}\;\;\;a_{{\vec k},\nu}(\tau)=
\tau^{-1}{\stackrel{\bullet} {\sf A}}_{{\vec k},\nu}(\tau),
\label{eq:A11}\end{eqnarray}  
which are directly connected to the ``physical'' components of the
electric field, ${\cal E}^m = \sqrt{-{\rm g}}{\rm g}^{ml}{\stackrel{\bullet} A}_l$~:
\begin{eqnarray}
\partial_\tau [\partial_{\tau}^{2}+{1\over \tau}\partial_{\tau}+
{\nu^2 \over \tau^2} +k_{\bot}^{2}]\varphi_{{\vec k},\nu}(\tau)=0~~,
\label{eq:A12}\end{eqnarray}                 
\begin{eqnarray}
\partial_\tau [\tau^2 \partial_{\tau}^{2}+\tau\partial_{\tau}+
\nu^2  +k_{\bot}^{2}\tau^2 ] a_{{\vec k},\nu}(\tau)=0~~,
\label{eq:A13}\end{eqnarray} 
As a result, we obtain that  the functions $\varphi_{{\vec k},\nu}(\tau)$
and $a_{{\vec k},\nu}(\tau)$ obey inhomogeneous Bessel equations,
\begin{eqnarray}
[\partial_{\tau}^{2}+{1\over \tau}\partial_{\tau}+
{\nu^2 \over \tau^2} + k_{\bot}^{2}]\varphi_{{\vec k},\nu}(\tau)
={\vec k}^{2}c_\varphi ~~,
\label{eq:A14}\end{eqnarray}             
\begin{eqnarray}
[\partial_{\tau}^{2}+{1\over \tau}\partial_{\tau}+
{\nu^2 \over \tau^2} +k_{\bot}^{2}] a_{{\vec k},\nu}(\tau)
=\tau^{-2} c_a ~~,              
\label{eq:A15}\end{eqnarray}             
where $c_\varphi$ and $c_a$  are arbitrary constants.
Now we may cast the solution of these equations in the
form of the sum of the partial solution of the inhomogeneous equation
and a general solution of the homogeneous equation,
\begin{eqnarray}
\Psi_{{\vec k},\nu}(\tau) = a  H^{(2)}_{-i\nu}(k_{\bot}\tau) +
a^*  H^{(1)}_{-i\nu}(k_{\bot}\tau)~~,
\label{eq:A16a}\end{eqnarray}       
\begin{eqnarray}
\varphi_{{\vec k},\nu}(\tau) = c  H^{(2)}_{-i\nu}(k_{\bot}\tau) +
c^*  H^{(1)}_{-i\nu}(k_{\bot}\tau) + c_\varphi s_{1,i\nu}(k_{\bot}\tau)~~,
\label{eq:A16}\end{eqnarray}             
\begin{eqnarray}
a_{{\vec k},\nu}(\tau) = \gamma  H^{(2)}_{-i\nu}(k_{\bot}\tau) +
\gamma^*  H^{(1)}_{-i\nu}(k_{\bot}\tau) + c_a s_{-1,i\nu}(k_{\bot}\tau)~~.
\label{eq:A17}\end{eqnarray}             
where $s_{\mu,\nu}(x)$ is the so-called Lommel function \cite{TF,Wat}. 

Furthermore, it is useful to notice that  the system of the Maxwell equations 
(\ref{eq:E2.10})-(\ref{eq:E2.11}) also has an infinite set of the
$\tau$-independent solutions of the form
\begin{eqnarray}
 W_i(\eta,{\vec r}) =\partial_i \chi (\eta,{\vec r})~~,
\label{eq:A19}\end{eqnarray}    
where $\chi$ is an arbitrary function of the spacial coordinates 
$\eta$ and ${\vec r}$.
Thus, they are the pure gauge solutions of the Abelian theory, compatible
with the gauge condition.     

In order to find the coefficients one should integrate Eqs. (\ref{eq:A16})
and (\ref{eq:A17}) with respect to the Hamiltonian time $\tau$, thus
finding the functions $\Phi$ and ${\sf A}$. Next, it is necessary to solve 
Eqs.(\ref{eq:A6}) for the Fourier components of the vector potential and to
substitute them into the original system of 
Eqs.~(\ref{eq:A2})--(\ref{eq:A5}).  Using functional relations from 
Appendix 2, one obtains that ~$c+\nu\gamma=0$ ~and ~$c_a-\nu c_\varphi=0$~.

One of the solutions, (already normalized according to Eq.(\ref{eq:E2.14}))
is found immediately: 
\begin{eqnarray}  
V^{(1)}_{{\vec k},\nu}(x)={e^{-\pi\nu/2}\over 2^{5/2}\pi k_{\bot}} 
\left( \begin{array}{c} 
                         k_y \\ 
                        -k_x \\ 
                         0 
                             \end{array} \right)
H^{(2)}_{-i\nu} (k_{\bot}\tau) e^{i\nu\eta +i{\vec k}{\vec r}}~~.
\label{eq:A18} 
\end{eqnarray} 
Initially, the components of the vector mode $V^{(2)}$
appear in the following form required by the convergence of the integral,
\begin{eqnarray}
 \left( \begin{array}{c}
            \nu  k_r R^{(2)}_{-1,-i\nu}(k_{\bot}\tau |S) \\
          -R^{(2)}_{1,-i\nu}(k_{\bot}\tau | s) -
                 i \nu [e^{\pi\nu /2}/ \sinh(\pi\nu /2)]
                               \end{array} \right)
e^{i\nu\eta +i{\vec k}{\vec r}}~~, \nonumber
\end{eqnarray}                                     
However, it can be gauge transformed to the more compact form,
\begin{eqnarray}  
V^{(2)}_{{\vec k},\nu}(x)={e^{-\pi\nu/2}\over 2^{5/2}\pi k_{\bot}} 
\left( \begin{array}{c} 
       k_r \nu R^{(2)}_{-1,-i\nu}(k_{\bot}\tau |s)  \\ 
       - R^{(2)}_{1,-i\nu}(k_{\bot}\tau | s) 
                                            \end{array} \right)
                e^{i\nu\eta +i{\vec k}{\vec r}}~~, 
\label{eq:A25}\end{eqnarray} 
The third solution, the last one by the count of the nonvanishing
components of the vector potential in the gauge $A^\tau =0$,
has the following form,   
\begin{eqnarray} 
V^{(3)}_{{\vec k},\nu}(x)=
    \left( \begin{array}{c} 
                 k_r  Q_{-1,i\nu}(k_{\bot}\tau) \\ 
                 \nu Q_{1,i\nu}(k_{\bot}\tau)
                                            \end{array} \right)
                   e^{i\nu\eta +i{\vec k}{\vec r}}~~.    
\label{eq:A26}\end{eqnarray}         

The modes $V^{(1)}$ and $V^{(2)}$ are the normalized solutions of the 
Maxwell equations. They are orthogonal and obey the normalization condition,
\begin{eqnarray} 
(V^{(1,2)}_{{\vec k},\nu},V^{(1,2)}_{{\vec k}',\nu'})=\delta(\nu-\nu')
\delta({\vec k}-{\vec k}'),\;\;\;
(V^{(1)}_{{\vec k},\nu},V^{(2)}_{{\vec k},\nu})=0~~.
\label{eq:A28}\end{eqnarray}  
which can be easily verified by means of Eq.~(\ref{eq:A2.5}). 
The norm of these solutions is given by the Eq.~(\ref{eq:E2.14}).         
A normalization coefficient of the mode $V^{(3)}$ 
is not defined as this mode has a zero norm. It is also
orthogonal to $V^{(1)}$ and  $V^{(2)}$:
\begin{eqnarray} 
(V^{(3)}_{{\vec k},\nu},V^{(3)}_{{\vec k}',\nu'})=
(V^{(1)}_{{\vec k},\nu},V^{(3)}_{{\vec k},\nu})=
(V^{(2)}_{{\vec k},\nu},V^{(3)}_{{\vec k},\nu})=0~~.
\label{eq:A30}\end{eqnarray}   
Thus this mode drops out from the decomposition of the free gauge field. 
 
The conservation of the constraint can be obtained as a consequence of the 
Eqs. ~(\ref{eq:A12}) and (\ref{eq:A13}) in the form,
\begin{eqnarray}
\tau \partial_\tau [\varphi_{{\vec k},\nu}(\tau)+
\nu a_{{\vec k},\nu}(\tau)] 
\equiv \tau\partial_\tau {\cal C}_{{\vec k},\nu}(\tau)=0~~,
\label{eq:A31}\end{eqnarray} 
which reassures us in consistency between the dynamic equations and
conservation of the Gauss law constraint.
  
One can explicitly check that the modes $V^{(1)}$ and $V^{(2)}$  obey
the constraint equation (\ref{eq:A10}), which expresses the Gauss law.
The mode $V^{(3)}$ does not. This mode corresponds to the longitudinal
field which cannot exist without the source.

\renewcommand{\theequation}{A2.\arabic{equation}}
\setcounter{equation}{0}  

\bigskip

\centerline {\bf APPENDIX 2. Mathematical miscellany}       

\bigskip

This appendix contains a list of mathematical formulae for the functions
which appear in various calculations in the body of the paper and Appendix~1.
The components of the vector field are expressed via  two types of
integrals. The first of them was studied in Ref.\cite{Wat}:
\begin{eqnarray}   
R^{(j)}_{\mu,\nu}(x |{\sf S})=\int x^\mu H^{(j)}_{\nu}(x) d x =
x[(\mu +\nu -1)H^{(j)}_{\nu}(x) {\sf S}_{\mu -1,\nu -1}(x)-
H^{(j)}_{\nu -1}(x){\sf S}_{\mu ,\nu}(x)]~~,
\label{eq:A2.1}\end{eqnarray}    
where ${\sf S}_{\mu ,\nu}$ stands for any of the two Lommel functions, 
$s_{\mu ,\nu}$ or $S_{\mu ,\nu}$ ~\cite{TF,Wat}. [ Whenever we omit the 
indicator $|{\sf S})$, the function $R^{(j)}_{\mu ,\nu}(x |s)$ is 
assumed.]  The second type of  integrals,  
\begin{eqnarray}   
Q_{\mu,\nu}(x) = \int_{0}^{x} x^\mu  d x s_{-\mu,\nu}(x). 
\label{eq:A2.2}\end{eqnarray} 
is a new one.
The functions $R^{(j)}_{\mu,\nu}(x|{\sf S})$  are introduced 
as the indefinite integrals. The preliminary choice of the lower limit
and, consequently, between $s_{\mu ,\nu}$ and $S_{\mu ,\nu}$,
is motivated by the requirement of the convergence and regular behavior.
One can easily prove that 
\begin{eqnarray}   
R^{{2 \choose 1}}_{-1,\mp i\nu}(x |S)-R^{{2 \choose 1}}_{-1,\mp i\nu}(x |s) = 
 {\mp i e^{\pi\nu /2} \over \nu \sinh(\pi\nu /2)}, \nonumber \\
R^{{2 \choose 1}}_{1,\mp i\nu}(x |S)-R^{{2 \choose 1}}_{1,\mp i\nu}(x |s) = 
 {\pm i \nu e^{\pi\nu /2} \over \sinh(\pi\nu /2)}.
\label{eq:A2.3}\end{eqnarray}  
We often use the following relation between the Lommel functions 
\cite{TF,Wat}
\begin{eqnarray}
{\sf S}_{1,i\nu}(k_{\bot}\tau)=1- \nu^2 {\sf S}_{-1,i\nu}(k_{\bot}\tau)~~.     
\label{eq:A2.4}\end{eqnarray} 
From the integral representations (\ref{eq:A2.1}) and  (\ref{eq:A2.2}),
it is straightforward to derive the functional relations
\begin{eqnarray}
R^{(j)}_{-1,i\nu}(k_{\bot} \tau)+ 
{1 \over \nu^2}R^{(j)}_{1,i\nu}(k_{\bot} \tau) 
= - {\tau \over \nu^2}{\partial \over \partial \tau}
H^{(j)}_{i\nu}(k_{\bot}\tau)~~,
\label{eq:A2.5}\end{eqnarray}
\begin{eqnarray}
Q_{-1,i\nu}(k_{\bot} \tau)-Q_{1,i\nu}(k_{\bot} \tau)=
-{\tau \over \nu^2} {\partial\over\partial\tau} s_{1,i\nu}(k_{\bot}\tau)=
\tau {\partial\over\partial\tau} s_{-1,i\nu}(k_{\bot}\tau)~~.
\label{eq:A2.6}\end{eqnarray} 
The Wronskian of the Hankel and Lommel functions,
\begin{eqnarray}   
W\{s_{1 ,i\nu}(x), H^{(j)}_{i\nu}(x)\} =
 - {1 \over x}  R^{(j)}_{1,i\nu}(x)~~,
\label{eq:A2.7}\end{eqnarray}       
is necessary to prove orthogonality of $V^{(2)}$ and $V^{(3)}$. To prove
(\ref{eq:A2.7}), one should use the following representation for the Lommel
function,
\begin{eqnarray}   
s_{1 ,i\nu}(x)= {\pi\over 4i}[ H^{(1)}_{i\nu}(x)R^{(2)}_{1,i\nu}(x)
- H^{(2)}_{i\nu}(x)R^{(1)}_{1,i\nu}(x)]~~,
\label{eq:A2.8}\end{eqnarray}       
which follows from Eq.~(\ref{eq:A2.1}) and its consequence,
\begin{eqnarray}   
s'_{1 ,i\nu}(x)= {\pi\over 4i}[ H^{(1)'}_{i\nu}(x)R^{(2)}_{1,i\nu}(x)
- H^{(2)'}_{i\nu}(x)R^{(1)}_{1,i\nu}(x)]~~.
\label{eq:A2.9}\end{eqnarray}   
In order to prove relation (\ref{eq:E2.30}) one should use representation
(\ref{eq:A2.1}) for the functions $R^{(j)}_{\nu,\nu}$ and the Wronskian
of two independent Hankel functions. The proof of relation (\ref{eq:E2.41})
begins with replacing the functions $R^{(j)}_{-1,i\nu}$ by  
$R^{(j)}_{1,i\nu}$ by means of Eq.~(\ref{eq:A2.5}). The final result 
follows from Eq.~(\ref{eq:A2.9}) and (\ref{eq:A2.6}).

\bigskip

\renewcommand{\theequation}{A3.\arabic{equation}}
\setcounter{equation}{0}  

\centerline {\bf APPENDIX~3. Calculation of the longitudinal part
            of the propagator} 

\bigskip
      
The kernels (\ref{eq:E2.42}) and (\ref{eq:E2.44}) of the longitudinal
and instantaneous parts of the propagator are given in the form of the
three--dimensional Fourier integrals $d\nu d^2{\vec k}$. Here, we describe
major steps of calculations which lead to Eqs.~(\ref{eq:E5.1})  and
(\ref{eq:E5.2}).

We permanently use the following integral representation for the Hankel 
functions,
\begin{eqnarray}
e^{-\pi\nu/2}e^{\pm i\nu\eta} H^{{2 \choose 1}}_{\mp i\nu}
(k_{\bot}\tau)= {\pm i\over \pi} \int_{-\infty}^{\infty} 
e^{\mp ik_{\bot}\tau \cosh (\theta-\eta)} e^{\pm i\nu\theta} d \theta~~,
\label{eq:A3.1}
\end{eqnarray}         
which allows one to calculate many integrals by changing the order
of integration.
The Lommel function $S_{1,i\nu}$ has a similar representation,
\begin{eqnarray} 
S_{1,i\nu}(x)= x\int_{0}^{\infty} \cosh u \cos\nu u ~e^{-x\sinh u} d u~~,
\label{eq:A3.2}\end{eqnarray}  
Integrating it by parts, and using Eq.~(\ref{eq:A2.4}), we find the
integral representation for $S_{-1,i\nu}$, 
\begin{eqnarray} 
\nu S_{-1,i\nu}(x)= \int_{0}^{\infty}  \sin (\nu u) ~e^{-x\sinh u} d u~~,
\label{eq:A3.3}\end{eqnarray}  
We start with integral representation (\ref{eq:A2.2}) of the functions 
$Q_{\pm 1,i\nu}$ and perform integration over $\nu$.     
To compute the integrals from the function $s_{1,i\nu}$ it can be
conveniently decomposed in the following way,
\begin{eqnarray} 
s_{1,i\nu}(x)= S_{1,i\nu}(x)-h_{i\nu}(x)~~,~~~ 
h_{i\nu}(x)={e^{-\pi\nu /2}\over 2}{\pi\nu /2 \over\sinh(\pi\nu /2 )} 
[H^{(1)}_{i\nu}(x)+ H^{(2)}_{-i\nu}(x)]~~.
\label{eq:A3.4}\end{eqnarray}        
which allows one to find 
\begin{eqnarray} 
\int_{-\infty}^{\infty}S_{1,i\nu}(k_{\bot}\tau)e^{i\nu\eta}d\nu= 
\pi k_{\bot}\tau\cosh\eta e^{-k_{\bot}\tau\sinh |\eta|}~~,\nonumber\\ 
\int_{-\infty}^{\infty} \nu S_{-1,i\nu}(k_{\bot}\tau) e^{i\nu\eta}d\nu= 
i\pi {\rm sign}\eta e^{-k_{\bot}\tau\sinh |\eta|}~~, 
\label{eq:A3.5}\end{eqnarray}  
The similar Fourier integrals from the function $h_{i\nu}$ are calculated
using the representation (\ref{eq:A3.1}) for the Hankel functions and 
and the integral,
\begin{eqnarray} 
{\pi\nu/2 \over \sinh(\pi\nu/2 ) }= {1\over 2}
\int_{-\infty}^{\infty} d\theta{e^{i\nu\theta}\over \cosh^2\theta}~~.
\label{eq:A3.6}\end{eqnarray}  
This yields, for example, 
\begin{eqnarray}
 \int_{-\infty}^{\infty} d\nu e^{i\nu\eta}h_{i\nu}(k_{\bot}\tau) =
\int_{-\infty}^{\infty} {d~\theta \over \cosh^2\theta}
\sin [k_{\bot}\tau\cosh(\theta-\eta)]~~.               
\label{eq:A3.7}\end{eqnarray}  
After integration over $\nu$ we obtain the following integral for
the $\Delta^{(L)}_{rs}$ 
\begin{eqnarray}
 \Delta^{(L)}_{rs}= \int {d^2{\vec k}\over (2\pi)^3}
{k_rk_s\over  k_{\bot}^2} e^{i{\vec k}{\vec r}}     
\int_{0}^{\tau_2}{d\tau\over\tau}
\bigg( \pi k_{\bot}\tau\cosh\eta e^{-k_{\bot}\tau\sinh |\eta|}
- \int_{-\infty}^{\infty} {d~\theta \over \cosh^2\theta}
\sin [k_{\bot}\tau\cosh(\theta-\eta)]\bigg)~~.               
\label{eq:A3.8}\end{eqnarray}  
and similar integrals for the other components. 
The first term in this formula is calculated in the following way.
After integration over $\tau$ we continue: 
\begin{eqnarray} 
 \Delta'_{rs}= - {\partial_r\partial_s \over 8 \pi^2} \coth |\eta|~ 
 \int {d^2{\vec k}\over  k_{\bot}^2} e^{i{\vec k}{\vec r}}     
[1- e^{-k_{\bot}\tau_2\sinh |\eta|}] \nonumber\\
= - {\partial_r\partial_s \over 4 \pi}  \coth |\eta|   
\int_{0}^{\infty} {d k_{\bot}\over k_{\bot}}  J_0(k_{\bot}r_{\bot})
[1-e^{-k_{\bot}\tau_2\sinh |\eta|}]=
- {\partial_r\partial_s \over 4 \pi}  \coth |\eta|  
\ln\bigg[ { \tau_2\sinh |\eta| + \sqrt{ r_{\bot}^2+\tau_{2}^{2}\sinh^2\eta }
\over r_{\bot}}\bigg]~~.                           
\label{eq:A3.9}\end{eqnarray}                             
To work out he second term, one should introduce $k_z=k_{\bot}\sinh\theta$
and  $k_0=k_{\bot}\cosh\theta=|{\bf k}|$ and join $d^2{\vec k}d\theta$ in
one three-dimensional integration $d^3{\bf k}$. 
With $t=\tau\cosh\eta$, ${\bf r}=(x,y,\tau\sinh\eta)$, this leads to
\begin{eqnarray} 
 \Delta''_{rs}=  {\partial_r\partial_s \over {2 \pi}^3} 
\int_{0}^{\tau_2}{d\tau\over\tau}
 \int {d^3{\bf k}\over  k_{0}^{3}} e^{i{\bf k}{\bf r}}\sin k_0 t
={\partial_r\partial_s \over 4\pi} \int_{0}^{\tau_2}{d\tau\over\tau}
\bigg( \theta(r_{\bot}^2-\tau^2)
{\tau\cosh\eta\over\sqrt{r_{\bot}^2+\tau^{2}\sinh^2\eta }} 
+ \theta(\tau^2-r_{\bot}^2) \bigg) \nonumber\\
= {\partial_r\partial_s \over 4\pi}
\bigg(\theta(r_{\bot}-\tau_2)~ \coth |\eta|~  
\ln\bigg[ { \tau_2\sinh |\eta| + \sqrt{ r_{\bot}^2+\tau_{2}^{2}\sinh^2\eta }
\over r_{\bot}}\bigg]  +\theta(\tau_2-r_{\bot})~
\ln {\tau_2\over r_{\bot}}\bigg)~~.        
\label{eq:A3.10}\end{eqnarray}                             
Adding (\ref{eq:A3.9}) and (\ref{eq:A3.10}), we obtain the first of the 
equations (\ref{eq:E5.1}).

\end{document}